\documentclass[11pt, a4paper]{article}
\pdfoutput=1
\usepackage[utf8]{inputenc}
\usepackage{amsmath}
\usepackage{mathtools}
\usepackage{braket}
\usepackage{bbold}
\usepackage{amsfonts}
\usepackage{amssymb}
\usepackage{geometry}
\usepackage{csquotes}
\usepackage{slashed}
\usepackage{tensor}
\usepackage{fancyhdr}
\usepackage{comment}
\usepackage{wrapfig}
\usepackage[toc,page]{appendix}
\usepackage[english]{babel}
\usepackage{hyperref}
\usepackage{graphicx}
\usepackage[export]{adjustbox}
\usepackage[disable]{todonotes}
\usepackage{xcolor}
\usepackage[skip=5pt]{caption}
\usepackage{subcaption}
\usepackage{authblk}
\usepackage[
backend=biber,sorting=none,doi=false,isbn=false,url=false,maxnames=10
]{biblatex}
\graphicspath{ {./kasner_figures/} }
\begin{document}

\begin{flushright}
QMUL-PH-21-60\\
\end{flushright}

\bigskip
\bigskip 
\bigskip

\centerline{\Large  Kasner geometries inside holographic superconductors}

\bigskip
\bigskip

\centerline{\bf Lewis Sword and David Vegh}

\bigskip

\begin{center}

\small{
{ \it   Centre for Theoretical Physics, Department of Physics and Astronomy \\
Queen Mary University of London, 327 Mile End Road, London E1 4NS, UK}}

\medskip

{\it Email:} \texttt{l.sword@qmul.ac.uk}, \texttt{d.vegh@qmul.ac.uk}

\end{center}

\bigskip 
\bigskip
\centerline{ \it \today}

\begin{abstract}
The recent study of holographic superconductors has shown the emergence of a Kasner universe behind the event horizon. This paper serves to add to the discussion by introducing two modifications to the holographic superconductor model: an axion field term and an Einstein-Maxwell-scalar (EMS) coupling term. We first discuss the effect the modification parameters have on the condensate then explore the black hole interior dynamics. Features previously identified in the interior are found in the model presented, including the collapse of the Einstein-Rosen bridge, Josephson oscillations and Kasner inversions/transitions. However, we find that by increasing the EMS coupling parameter, the collapse does not occur near the axion-Reissner-Nordstr\"om horizon and the oscillations are no longer present; the geometry entering into a Kasner regime after a large-$r$ collapse instead. 
\end{abstract}

\tableofcontents

\newpage

\section{Introduction}

The original framework for holographic superconductors was presented in 2008 \cite{Hartnoll:2008vx,Hartnoll:2008kx} based on the previous work of Gubser \cite{Gubser:2008px}. These papers detailed how, via the AdS/CFT correspondence \cite{Maldacena:1997re,Witten:1998qj,Gubser:1998bc}, a superconductor can be mapped to its gravitational dual by introducing a black hole and charged scalar field, providing a temperature and condensate respectively. Specifically, \cite{Hartnoll:2008vx} described the ``probe limit", whereby taking the charge of the scalar field to infinity, the back-reaction effects can be deemed negligible while \cite{Hartnoll:2008kx} extended past this limit, producing results for various finite charges.

Attention has recently turned to numerically investigating behind the horizon of black holes \cite{Frenkel:2020ysx}. Here it was found that the geometry near to the singularity is a Kasner universe \cite{Kasner:1921zz}, indicating a flow from anti-de Sitter spacetime. Following this, \cite{Hartnoll:2020fhc} showed that the interior of a holographic superconductor divides into specific epochs; collapse of the Einstein-Rosen (ER) bridge, Josephson oscillations of the scalar field, and Kasner geometries with potential Kasner inversions. These ideas have been shown to persist in other models. In \cite{Mansoori:2021wxf}, they introduce a massless scalar field (named the ``axion" field \cite{Andrade:2014xca} due to the shift symmetry) which serves to break translational invariance in the boundary field theory\footnote{The notion of breaking translational invariance follows from the initial work of \cite{Vegh:2013sk,Andrade:2013gsa}.} and identify the ER bridge collapse and Kasner geometry. Also, following from the work on scalarised black holes by including Einstein-Maxwell-scalar (EMS) terms \cite{Herdeiro:2018wub,Fernandes:2019rez}, even asymptotically flat spacetimes have been shown to exhibit the mentioned dynamical epochs \cite{Dias:2021afz}\footnote{The interior of an asymptotically flat, Einstein-Gauss-Bonnet black hole with scalar hair was studied in \cite{Grandi:2021ajl}, where the ER bridge collapse was also found.}.

This paper constructs an extension to the holographic superconductor model by introducing both an axion and EMS term to the action, then investigates their effects on phase transitions and the black hole interior. Initially, we explore the condensate results upon varying the coupling parameters of these terms. We find that for various electric charge values, increasing the EMS parameter leads to a smaller condensate, while increasing the axion parameter leads to larger values of the condensate, in agreement with \cite{Kim:2015dna}. We then focus on the black hole interior and show that by tuning the EMS parameter to a sufficiently large value, the Einstein-Rosen bridge collapse is not located near the axion-Reissner-Nordstr\"om (aRN) horizon and the Josephson oscillations are no longer present. Instead, a growth of the metric time component, $g_{tt}$, becomes apparent followed by a large-$r$ collapse, in contrast to behaviours when the EMS coupling is small. We argue that this is due to terms in the equations of motion becoming significant in the vicinity of the aRN horizon. These terms prevent the oscillatory form of the scalar field arising. From numerical inspection we identify that the geometry past the large-$r$ collapse point is Kasner. Additionally, it is shown that a large-$r$ Kasner geometry \cite{Frenkel:2020ysx} exists even in the presence of the axion. More specifically, the Kasner inversion behaviour still holds for non-zero axion parameter, showing that its effect on the large-$r$ geometry is negligible\footnote{This is the case for the main axion model considered, however we will discuss another where this no longer holds.}. This is a prominent feature of \cite{Hartnoll:2020fhc,Hartnoll:2020rwq} where contributions from the cosmological constant, mass and charge terms in the equations of motion are similarly neglected. Turning on both axion and EMS parameters leads to Kasner transitions as seen in \cite{Dias:2021afz}, rather than typical inversions.

The structure of the paper is as follows. Section \ref{sec:the_model} provides the model and framework, stating the field ans\"atze and the equations of motion on which the numerical analysis is based. Section \ref{sec:phase_diagrams} looks at the phase diagrams for a variety of parameter values, at charges $q = 1,3 \text{ and } 12$. Section \ref{sec:behind_the_horizon} investigates the superconductor interior, studying  the ER bridge collapse, the Kasner geometry and how parameter variation effects them. An alternative axion potential is considered in section \ref{sec:X_cubed_results} and finally, section \ref{s:discussion} summarises the findings and discusses potential future directions.


\section{The model}
\label{sec:the_model}

The model employed combines the original $(3+1)$-dimensional, $U(1)$ holographic superconductor action of \cite{Hartnoll:2008kx}, with an axion field \cite{Mansoori:2021wxf} and an Einstein-Maxwell-scalar (EMS) coupling term \cite{Herdeiro:2018wub,Dias:2021afz,Dias:2021vve}. The resulting action is $S = S_{1} + S_{2} + S_{3}$ with
\begin{align}
\label{eqn:TM_Action}
S_{1} &= \int d^4 x \sqrt{-g}   \left[ R  + \frac{6}{L^2} \right]\,, \\
S_{2} &= \int d^4 x \sqrt{-g}  \left[   - \frac{L^2}{4} F^2 - g^{\mu \nu}\left( \partial_{\mu} \psi - i q A_{\mu} \psi \right) \left( \partial_{\nu} \psi^{*} + i q A_{\nu} \psi^{*}\right) - m^2 |\psi|^2  \right]\,, \\
 S_{3} &=  \int d^4 x \sqrt{-g}  \left[ - \frac{L^2}{4} \gamma  F^2 |\psi|^2 - \frac{1}{L^2} K(X) \right] \,,
\end{align}

where $F = dA$, $A_{\mu}$ is the gauge field, $\psi$ is the charged complex scalar field with charge $q$ and mass $m$, $R$ is the Ricci scalar, $\gamma$ is the EMS coupling parameter, $L$ is the radius of anti de-Sitter (AdS) spacetime and $K(X)$ is the K-essence that introduces an axion field\footnote{The holographic axion model without the EMS term has been explored in depth in \cite{Baggioli:2015zoa}, which looks at various axion potentials.}. In this paper we take 
\begin{equation}
\label{eqn:TM_X}
X = \frac{L^2}{2}  \sum_{I} g^{\mu \nu} \partial_{\mu}\Omega^{I} \partial_{\nu} \Omega^{I}\,,
\end{equation}
where $\Omega^{I}$ is a massless scalar field which will be referred to as the axion, following the same terminology used in \cite{Mansoori:2021wxf,Andrade:2014xca,Baggioli:2021xuv}. We define it as
\begin{equation}
\label{eqn:TM_axion_field}
\Omega^{I} = \lambda_{0} x^{I}\,,
\end{equation}
with $I = (x, y)$ and axion parameter $\lambda_{0}$. Following previous convention, a gauge is chosen where the phase of the scalar field is constant, such that $\psi \in \mathbb{R}$. The radially dependent ans\"atze for the fields are
\begin{equation}
\label{eqn:TM_scalar_gauge_ansatz}
\psi = \psi^{*} = \psi(r)\, , \quad A = A_{0}(r)dt = \phi(r) dt\,,
\end{equation}
alongside metric ansatz
\begin{equation}
\label{eqn:TM_metric_ansatz}
ds^2 = \frac{L^2}{r^2} \left( -f(r) e^{-\chi(r)} dt^2 +\frac{1}{f(r)}dr^2 +  dx^2 + dy^2 \right) \,.
\end{equation}

The equations of motion that follow from the variation of the total action $S$ using these  ans\"atze are 
\begin{subequations}
\begin{align}
-\frac{2 q^2 \psi^2 \phi }{r^2 f \left(\gamma  \psi^2+1\right)}+\frac{\phi ' \left(\left(\gamma  \psi^2+1\right) \chi ' +4 \gamma  \psi \psi '\right)}{2 \left(\gamma  \psi^2 +1\right)}  + \phi '' &= 0\,, \label{eqn:TM_EOM1} 
\\
 \psi  \left(-\frac{L^2 m^2}{r^2 f} + \frac{q^2 e^{\chi} \phi^2}{f^2} + \frac{\gamma  r^2 e^{\chi} (\phi ')^2}{2 f}\right) + \psi ' \left(\frac{f'}{f}-\frac{\chi '}{2}-\frac{2}{r}\right) + \psi '' &=0 \,, \label{eqn:TM_EOM2}
\\
\frac{q^2  e^{\chi } \phi^2 \psi^2}{f^2} - \frac{\chi' }{r} +  (\psi ')^2 &= 0\,, \label{eqn:TM_EOM3}
\\
\frac{6}{r^2}-\frac{6}{r^2 f}-\frac{2 f'}{r f}+\frac{L^2 m^2 \psi^2}{r^2 f} +\frac{ K(X)}{r^2 f}+\frac{q^2 e^{\chi} \psi^2 \phi^2}{f^2}+ \frac{r^2 e^{\chi} \left(\gamma  \psi^2+1\right) (\phi ')^2}{2 f}+(\psi ')^2 &= 0\,, \label{eqn:TM_EOM4}
\end{align}
\end{subequations}

with \eqref{eqn:TM_EOM1} and \eqref{eqn:TM_EOM2} being the gauge field and scalar field equations respectively, while \eqref{eqn:TM_EOM3} and \eqref{eqn:TM_EOM4} are the independent metric field equations. In this paper we consider K-essence potentials of the form $K(X) = X^n$ and explicitly provide results for $K(X) = X$ in the main sections \ref{sec:phase_diagrams} and \ref{sec:behind_the_horizon}, while presenting a condensed discussion of the $K(X) = X^3$ model results in section \ref{sec:X_cubed_results}. The ans\"atze selected for the fields above lead to a simple form of $K(X)$ in terms of the radial coordinate, namely $K(X) = X^{n} = (r^2 \lambda_{0}^2)^n$.

The equations of motion abide by the following scaling symmetries:

\begin{itemize}
\item Time scaling,
\begin{equation}
\label{eqn:TM_chi_scaling_sym}
e^{\chi} \to a_{1}^2 e^{\chi}, \quad t \to a_{1} t, \quad \phi \to \frac{\phi}{a_{1}} \,.
\end{equation}

\item Horizon radius ($r_{h}$) scaling,
\begin{equation}
\label{eqn:TM_hor_scaling_sym}
r \to \frac{r}{a_{2}}, \quad \phi \to a_{2} \phi, \quad \lambda_{0} \to a_{2} \lambda_{0} \,.
\end{equation}

\item AdS radius ($L$) scaling,
\begin{equation}
\label{eqn:TM_ads_scaling_sym}
r \to \frac{r}{a_{3}}, \quad e^{\chi} \to a_{3}^2 e^{\chi}, \quad L \to a_{3} L, \quad m \to \frac{m}{a_{3}}, \quad \lambda_{0} \to a_{3} \lambda_{0} \,.
\end{equation}
\end{itemize}

Here $a_{i}, i = 1,2,3$ are scaling parameters. Equation \eqref{eqn:TM_chi_scaling_sym}
lets us take an arbitrary value of $\chi$ at the horizon, provided that $\chi \to 0$ at the UV boundary after rescaling. Symmetries \eqref{eqn:TM_hor_scaling_sym} and  \eqref{eqn:TM_ads_scaling_sym} are used to set $r_{h} = 1$ and $L = 1$, by taking $a_{2} = r_{h}$ and $a_{3} = 1/L$. These are the values applied throughout the numerical calculations.

Given the coupled nature of the equations of motion, we look to solve them numerically using Mathematica \cite{Mathematica_ref}. The general idea is to first produce series expansions of the fields both at the horizon, $r=r_{h}$, and the UV boundary, $r=0$, and then apply a shooting method\footnote{It is important to pick the ground state solution, where the scalar field exhibits no nodes.}. As is standard for the blackening factor, we have $f(r_{h}) = 0$, while to ensure that we have a finite norm of the gauge field, we require that $\phi(r_{h}) = 0$ also. Applying these conditions, the following horizon field expansions are constructed
\begin{subequations}
\begin{align}
\psi &= \psi_{h1} + \psi_{h2}(r-r_{h}) + \psi_{h3}(r-r_{h})^2 + \dots \label{eqn:TM_psi_hor} \\
f &= f_{h1} (r-r_{h}) + f_{h2} (r-r_{h})^2  + \dots \label{eqn:TM_f_hor} \\
\chi &= \chi_{h1} + \chi_{h2} (r-r_{h}) + \chi_{h3} (r-r_{h})^2 + \dots \label{eqn:TM_chi_hor} \\
\phi &= \phi_{h1}(r-r_{h}) + \phi_{h2} (r-r_{h})^2 +\phi_{h3} (r-r_{h})^3 + \dots  \label{eqn:TM_phi_hor} 
\end{align}
\end{subequations}
Inserting these into the equations of motion, series expanding, then equating the coefficients of each series to zero, we find the equations (\ref{eqn:TM_psi_hor})-(\ref{eqn:TM_phi_hor}) are fully determined by three parameters: $\phi_{h1}$, $\psi_{h1}$ and $\chi_{h1}$. As for the UV boundary, following the same procedure and setting $m^2 = -2/L^2$ which satisfies the Breitenlohner-Freedman bound\footnote{The general expansion of the scalar field at the UV boundary takes form $\psi =  \psi_{(1)}r^{d-\Delta} + \psi_{(2)}r^{\Delta} + \dots$, where $\Delta = \frac{d}{2}+ \sqrt{\frac{d^2}{4} +m^2}$: the dual boundary operator's ($\mathcal{O}$) conformal dimension. Taking $m^2 = -2/L^2$ then simplifies the expansion to equation \eqref{eqn:TM_psi_uv}.} \cite{Breitenlohner:1982jf}, the field expansions obtained are 
\begin{subequations}
\begin{align}
\psi &= \psi_{(1)} r +  \psi_{(2)} r^2 + \dots  \label{eqn:TM_psi_uv}\\
f &= 1 + f_{(1)}r^2 + f_{(2)} r^3 + \dots +  \tilde{f} r^{2n} + \dots \label{eqn:TM_f_uv} \\
\chi & = \chi_{(1)} + \frac{1}{2}\psi_{(1)}^{2} r^2 + \frac{4}{3} \psi_{(1)}\psi_{(2)} r^3 + \dots \label{eqn:TM_chi_uv} \\
\phi &= \phi_{(1)} + \phi_{(2)} r + q^2   \psi_{(1)}^2  \phi_{(1)} r^2  + \dots  \label{eqn:TM_phi_uv}
\end{align}
\end{subequations}

As in \cite{Hartnoll:2008vx,Hartnoll:2008kx} the UV boundary expansion \eqref{eqn:TM_phi_uv} determines the chemical potential $\mu$ and the charge density $\rho$
\begin{equation}
\label{eqn:TM_chem_charge_read_off}
\phi_{(1)} = \mu \,, \quad \phi_{(2)} = - \rho \,.
\end{equation}
There are also UV boundary conditions on the scalar field $\psi$. Depending on choice of quantisation \cite{Klebanov:1999tb}, either
\begin{align}
\psi_{(1)} &= 0 \quad \text{and} \quad \psi_{(2)} = \frac{1}{\sqrt{2}} \langle \mathcal{O} \rangle \,, \quad \text{or} \label{eqn:TM_bc_1} \\
\psi_{(2)} &= 0 \quad \text{and} \quad \psi_{(1)} =  \frac{1}{\sqrt{2}} \langle \mathcal{O} \rangle \,, \label{eqn:TM_bc_2} 
\end{align}
where $\langle \mathcal{O} \rangle$ is the condensate (the expectation value of the boundary operator dual to $\psi$). The free UV boundary parameters; $\phi_{(1)}$, $\phi_{(2)}$, $f_{(2)}$, $\chi_{(1)}$, and $\psi_{(2)}$ (or $\psi_{(1)}$) are then identified by employing a shooting method that satisfies condition \eqref{eqn:TM_bc_1} (or \eqref{eqn:TM_bc_2}). Note that the coefficients $f_{(1)}$ and $\tilde{f}$ in equation \eqref{eqn:TM_f_uv} can be written in terms of the aforementioned free UV parameters. Generally, $\tilde{f}$ takes form $\tilde{f} = (c_{1} \lambda_{0}^{2n} + \dots)$ with constant $c_{1}$, showing that for large $n$, the axion term's contribution to the UV boundary expansion becomes less important.

Having established the condensate, the temperature is now identified by passing to Euclidean signature of the metric and using the periodicity of the Euclidean time coordinate. In terms of the metric functions the temperature is then
\begin{equation}
\label{eqn:TM_temp_eqn}
T_{0} = \frac{|f'(r)|e^{-\chi(r)/2}}{4\pi} \bigg\rvert_{r = r_{h}} \,.
\end{equation}
Using the expansion parameters explicitly
\begin{equation}
\label{eqn:TM_temp_eqn_new}
T_{0} = \frac{ e^{-\frac{\chi(r_{h})}{2}}  }{16 \pi r_h} \left[ 2 (\lambda_{0}^2 r_{h}^2)^{n} - 12 +  e^{\chi(r_{h})} \left(\phi'(r_{h})\right)^2  r_{h}^4 + \left(2 L^2 m^2 + \gamma e^{\chi(r_{h})} \left(\phi'(r_{h})\right)^2 r_{h}^4  \right)\psi(r_{h})^2 \right] \,.
\end{equation}
The temperature is therefore completely determined by the fields at the horizon and more specifically, by the three horizon parameters identified in expansions \eqref{eqn:TM_psi_hor}, \eqref{eqn:TM_chi_hor}, \eqref{eqn:TM_phi_hor} since $\psi(r_{h}) = \psi_{h1}$, $\chi(r_{h}) = \chi_{h1}$ and $\phi'(r_{h}) = \phi_{h1}$. Note that in the numerical calculations, $T_{0}$ is rescaled by an overall factor of $e^{\chi_{(1)}/2}$ according to equation \eqref{eqn:TM_chi_scaling_sym}. 

\newpage
\section{Phase diagrams}
\label{sec:phase_diagrams}

The framework of the previous section allows us to check where normal phase crosses into superconducting phase by plotting the condensate as a function of temperature. Having set the scalar field mass previously, this leaves three parameters that potentially have effects on this relation: $q$, $\lambda_{0}$ and $\gamma$. In this section we make plots of fixed $q$ and focus on how the condensate behaves for different choices of the axion or EMS parameter, specifically for the $K(X) = X$ model. These plots use the boundary condition of \eqref{eqn:TM_bc_1} otherwise known as ``standard quantisation" and also feature dimensionless quantities 
\begin{equation}
\label{eqn:PD_dimensionless_quantities}
\sqrt{\langle \mathcal{\tilde{O}} \rangle} = \frac{\sqrt{\langle \mathcal{O} \rangle}}{\sqrt{\rho}} \,,  \quad T = \frac{T_{0}}
{\sqrt{\rho}}  \quad \text{ and } \quad \lambda = \frac{\lambda_{0}}{\sqrt{\rho}}\,. 
\end{equation}
where $\langle \mathcal{O} \rangle$ is defined in \eqref{eqn:TM_bc_1}. $T$ and $\lambda$ as defined in \eqref{eqn:PD_dimensionless_quantities} will also be used in the consequent sections. Finally, the arbitrary horizon value of $\chi$ is set to $\chi_{h1} = 1$.

\begin{figure}
\centering
\begin{subfigure}{1.0\textwidth}
	\centering 
	\includegraphics[width=13cm, height=7.2cm]{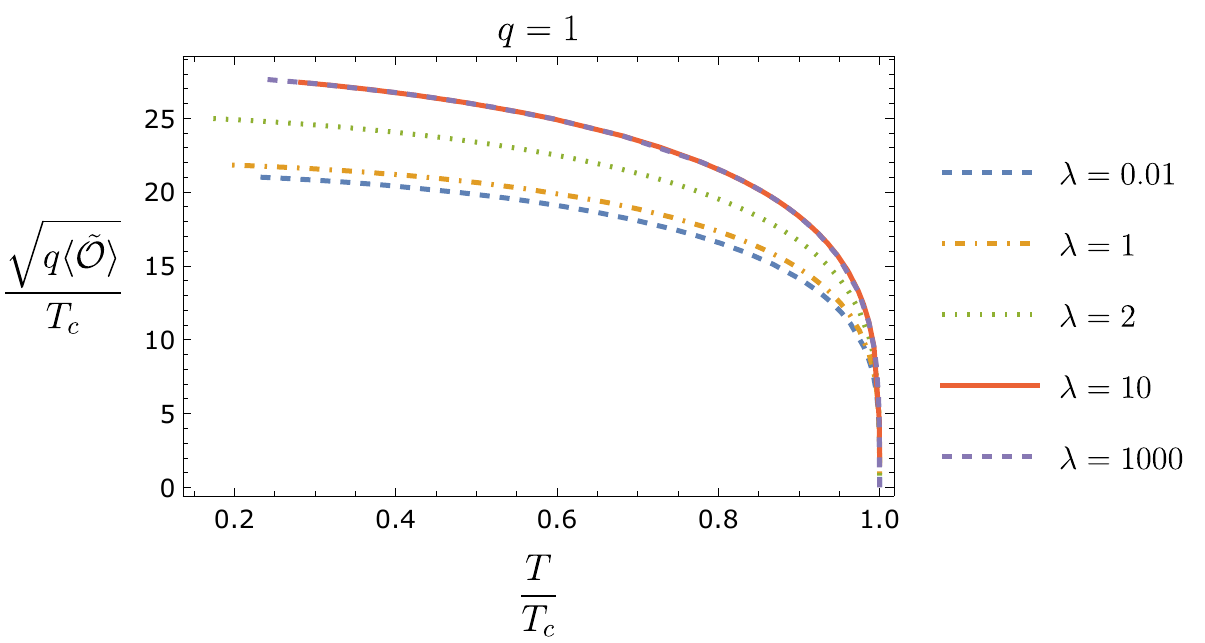}
	\vspace{0.3cm}
\end{subfigure}%

\begin{subfigure}{1.0\textwidth}
	\centering 
	\includegraphics[width=13cm, height=7.2cm]{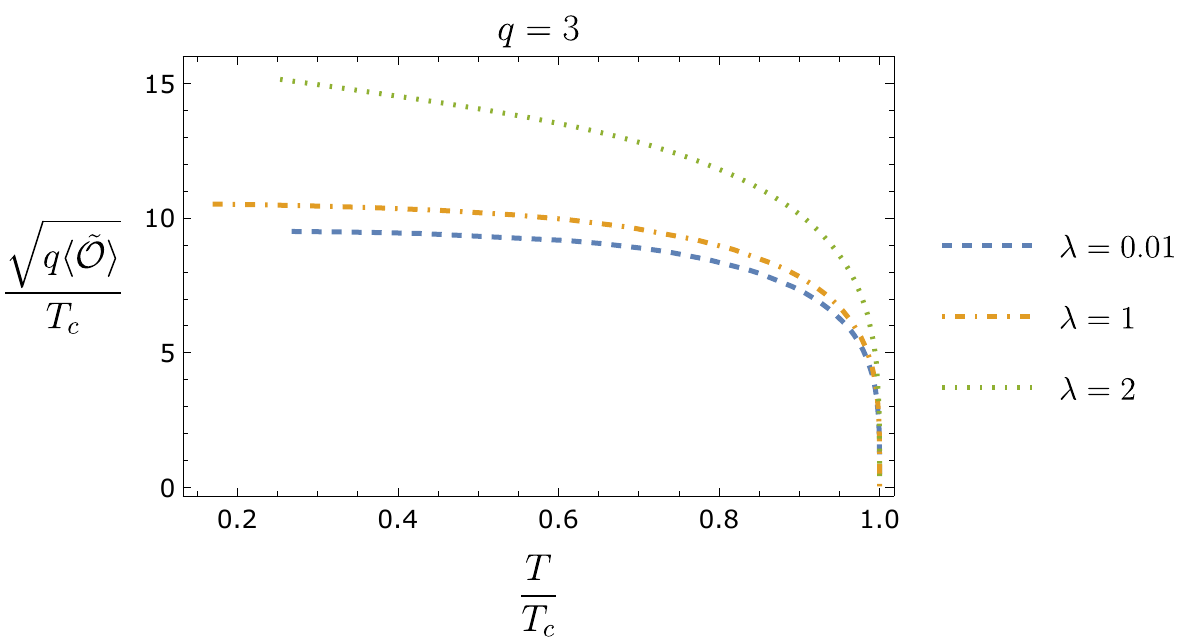}
	\vspace{0.3cm}
\end{subfigure}

\begin{subfigure}{1.0\textwidth}
	\centering 
	\includegraphics[width=13cm, height=7.2cm]{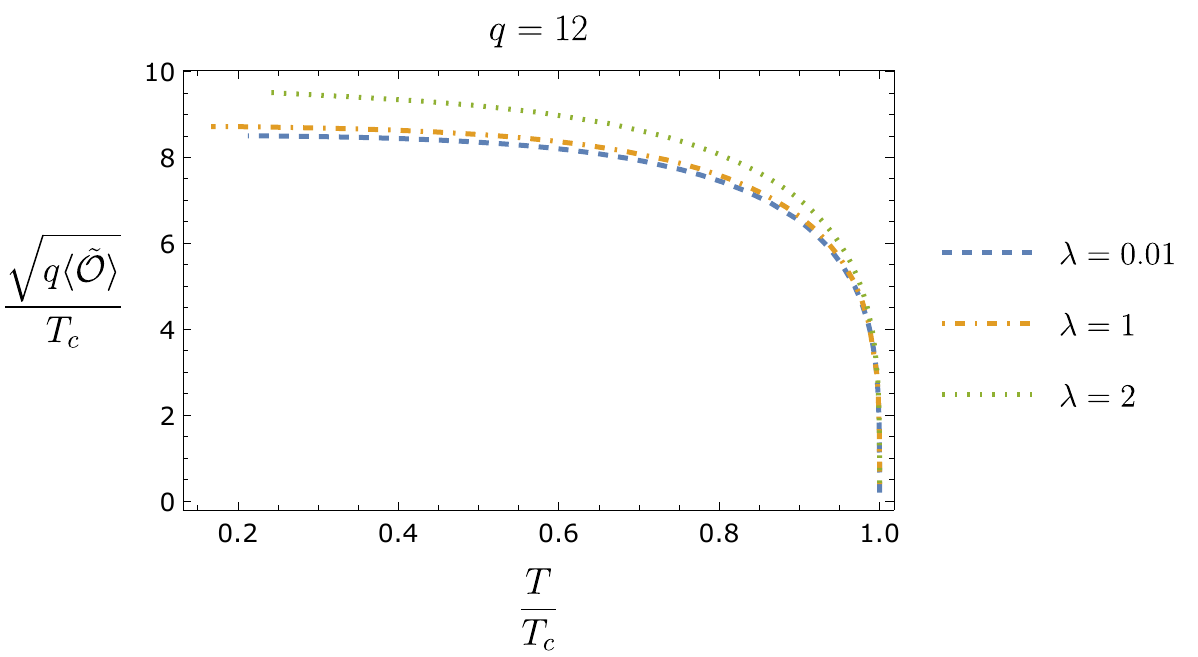}
\end{subfigure}
\caption{Condensate plots for $q=1$, $q=3$ and $q=12$ at $\gamma = 0$ and various $\lambda$. Increasing $\lambda$ has the effect of increasing the condensate to an apparent limiting point. This is demonstrated by the almost exact equivalence between the $\lambda = 10$ and $\lambda = 1000$ curves in the $q = 1$ plot. Increasing the charge generally pushes the condensate to its probe limit value. }
\label{fig:phase_diagram_lambda}
\end{figure}

Turning on $\lambda$ and turning off $\gamma$ allows us to inspect the effect of the axion contribution alone. The three plots of Figure \ref{fig:phase_diagram_lambda} correspond to charges $q = 1,3$ and $12$ and the curves within these plots present the condensate vs. temperature data at a fixed $\lambda = \lambda_{0}/ \sqrt{\rho}$. This is achieved using a double shooting method. By first choosing some $\lambda_{0}$, we trial initial horizon data, ($\phi_{h1}$, $\psi_{h1}$), to find which of these pairs satisfies \textit{both} condition \eqref{eqn:TM_bc_1} as well as $\lambda = \lambda_{\text{test}}$, where $\lambda_{\text{test}}$ is the value we have chosen to fix. The process is repeated for different $\lambda_{0}$ providing a suitable range of $T/T_{c}$ values. The key result is that by increasing $\lambda$, the value of the condensate increases, in agreement with the findings of \cite{Kim:2015dna}. There appears to be a convergence of the condensate value as we go to larger $\lambda$, demonstrated specifically by the coincidence of the  $\lambda = 10$ (red, solid) and $\lambda = 1000$ (purple, dashed) curves in the top $q=1$ plot. As $\lambda \to 0$ on the other hand, we find that at low temperatures the condensate approaches the same values stated in \cite{Hartnoll:2008kx} as expected, since turning both $\lambda$ and $\gamma$ parameters off, reduces the present model to the original. Also, an increase in charge $q$ clearly shows a reduction in the condensate value for a chosen $\lambda$.

\begin{figure}
\centering
\begin{subfigure}{1.0\textwidth}
	\centering 
	\includegraphics[width=13cm, height=7.2cm]{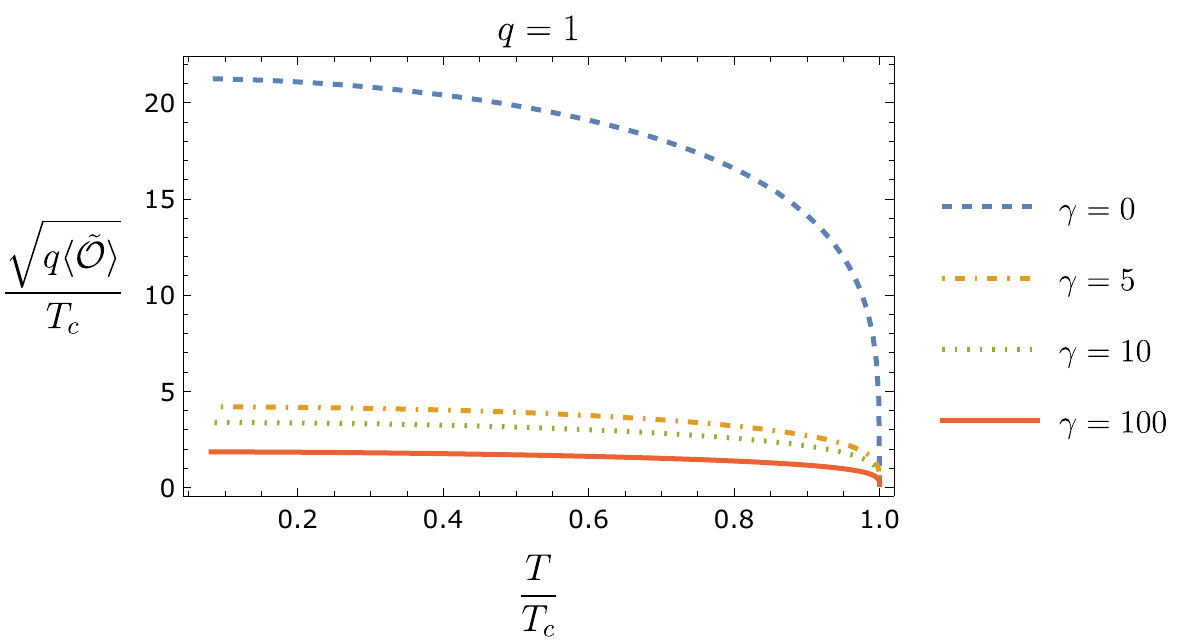}
	\vspace{0.3cm}
\end{subfigure}%

\begin{subfigure}{1.0\textwidth}
	\centering 
	\includegraphics[width=13cm, height=7.2cm]{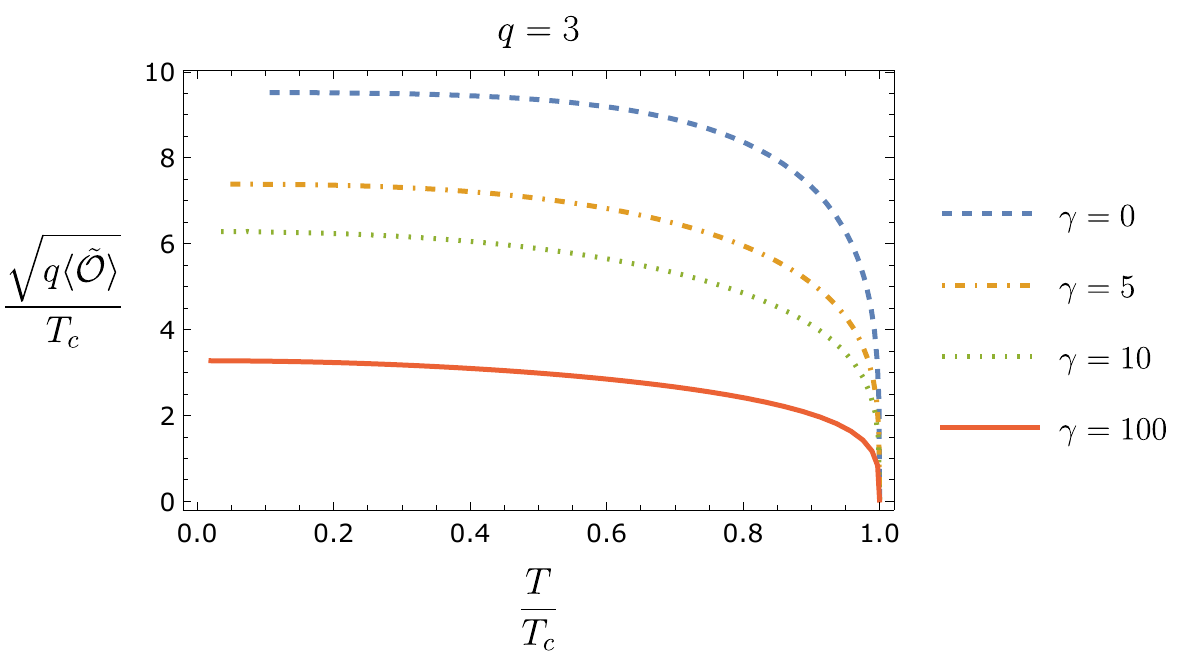}
	\vspace{0.3cm}
\end{subfigure}

\begin{subfigure}{1.0\textwidth}
	\centering 
	\includegraphics[width=13cm, height=7.2cm]{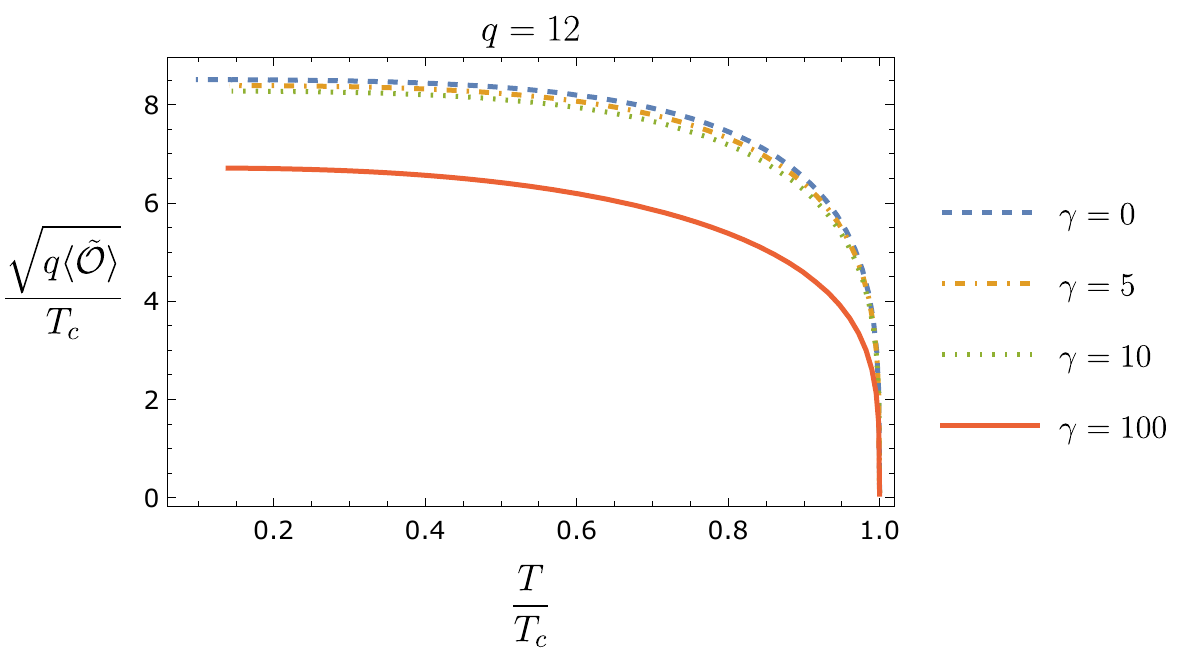}
\end{subfigure}
\caption{Condensate plots for $q = 1$, $q=3$ and $q=12$ at $\lambda = 0$ and various $\gamma$. Increasing $\gamma$ produces a decrease in the condensate for temperatures below $T_{c}$, while increasing the charge $q$ acts to constrict the condensate envelope, i.e. the difference in condensate value between the largest and smallest $\gamma$ is reduced.}
\label{fig:phase_diagram_gamma2}
\end{figure}

Given that $\gamma$ has no scaling in the symmetries mentioned, it is dimensionless and we can plot curves for its various values directly. Therefore, by turning on $\gamma$ and turning off $\lambda$, Figure \ref{fig:phase_diagram_gamma2} demonstrates the effects of varying $\gamma$ on the condensate, for charges $q =1,3$ and $12$. 
The main result observed is a reduction in the condensate as $\gamma$ is increased. As was the case for $\lambda \to 0$ in Figure \ref{fig:phase_diagram_lambda}, when $\gamma \to 0$ we once again retrieve the original holographic superconductor model results of \cite{Hartnoll:2008kx}. This can be seen by the blue dashed, $\gamma = 0$ curves. A natural conclusion from the alternative charge plots is that approaching larger $q$ acts to reduce the ``condensate envelope" which describes the values of the condensate between the largest and smallest $\gamma$ plotted. This is potentially related to the probe limit of \cite{Hartnoll:2008vx}, where when the charge terms in the equations of motion become more dominant, choice of $\gamma$ may become less relevant. For example, the $q=12$ plot shows a tighter grouping of curves, close to $\sqrt{q \langle \tilde{\mathcal{O}} \rangle} /T_{c} \approx 8.5$ which is exactly the probe limit of the original model. 

\newpage

\section{Inside the holographic superconductor}
\label{sec:behind_the_horizon}

Having studied the condensate, a feature of the black hole exterior, this section proceeds to investigate the interior dynamics of our modified holographic superconductor for the $K(X) = X$ model. Generally, the three main epochs established in \cite{Hartnoll:2020fhc}; collapse of the Einstein-Rosen (ER) bridge, Josephson oscillations and Kasner geometry, persist in this model, however, numerical observations based on parameter choice warrant further discussion. We show that when $\gamma$ takes sufficiently large values, there are no Josephson oscillations and the ER bridge collapse does not occur around the aRN horizon. Instead, there is first a growth phase of the bridge followed by a collapse at large-$r$.  We then briefly review the method in identifying the Kasner geometry in the large-$r$ Kasner epoch, based on the numerical field solutions. Using this setup, we investigate the Kasner geometry as well as both inversions \cite{Hartnoll:2020fhc} and transitions \cite{Dias:2021afz} under different choice of $\lambda$ and $\gamma$. We find that the axion term can be deemed negligible in the large-$r$ Kasner epoch based on its contribution to the equations of motion when evaluated on the original holographic superconductor solutions. This is then explicitly confirmed by numerical example using the Kasner inversion rule. Finally, we return to the large $\gamma$ behaviour and show that after the large-$r$ collapse, the interior geometry is well described by a Kasner universe.

\subsection{Collapse of the Einstein-Rosen bridge}
\label{ssec:er_bridge}

\begin{figure}[htb!]
\begin{center}
\includegraphics[width=8cm, height=6.5cm]{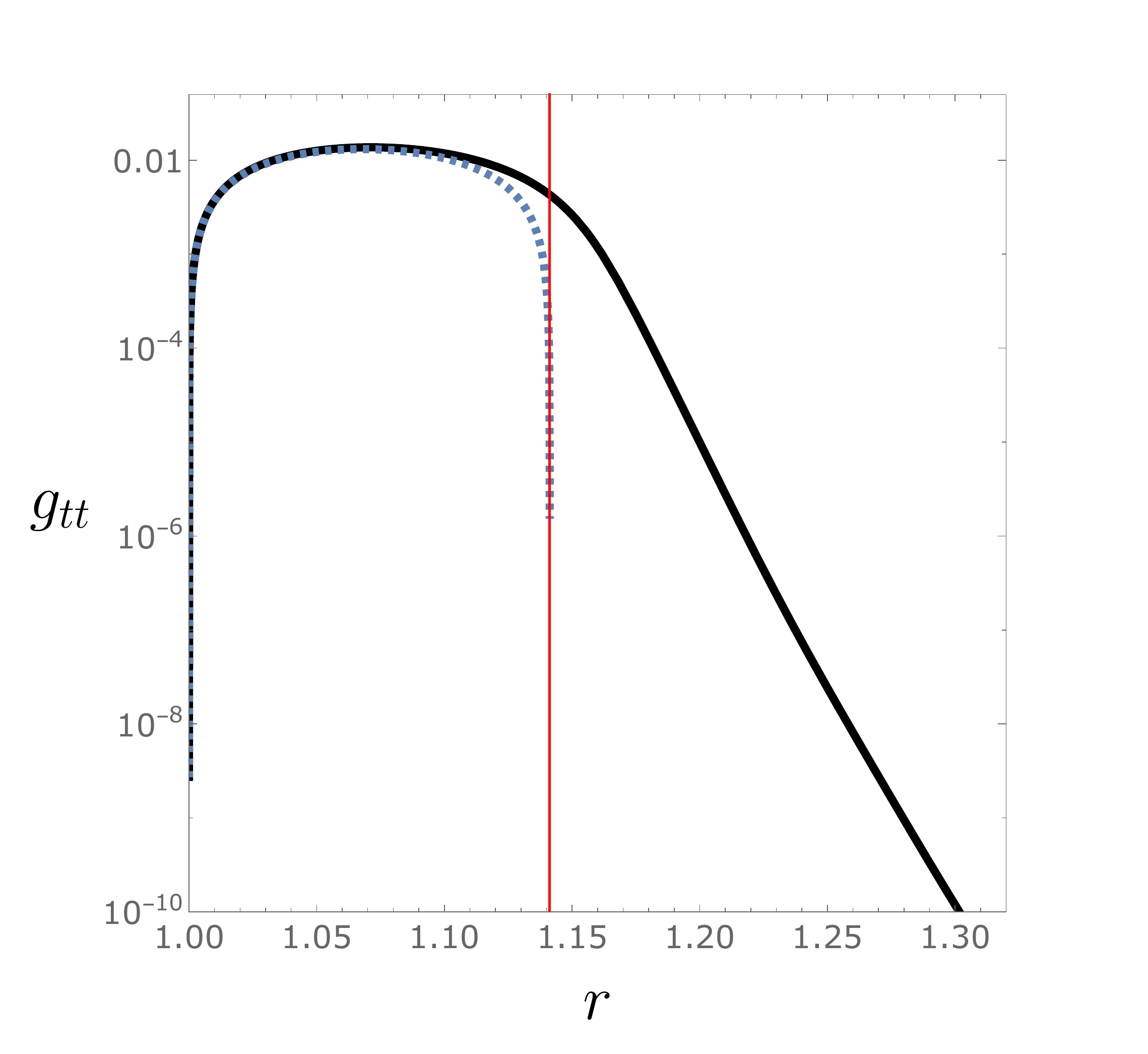}
\end{center}
\caption{The figure depicts two solutions of $g_{tt}$. The dashed blue curve is the aRN $g_{tt}$ solution (where $\psi = 0$) and it approaches zero exactly at $r_{\text{aRN}}$, depicted by the red vertical line. The black curve is the full numerical, $\psi \neq 0$ solution. Here, instead of approaching zero at $r_{\text{aRN}}$ it starts to exponentially decay - this is the ER bridge collapse. Here $q=1$, $\lambda = 2$, $\gamma =0$ and $T/T_{c} \approx 0.9954$.}
\label{fig:psizero_psinonzero_gtt}
\end{figure}

The collapse of the Einstein-Rosen bridge was established in \cite{Hartnoll:2020rwq}, describing the rapid decrease in the metric time component $g_{tt}$ over a short radial range, within the black hole. The collapse occurs at the ``would-be" inner horizon\footnote{``Would-be" inner horizon in the setting of \cite{Hartnoll:2020rwq} simply implies that if $\psi \neq 0$, there is no inner horizon. For discussion of the non-existence of inner horizons in the presence of scalar hair, see \cite{Cai:2020wrp} for general arguments.} of the metric solution when $\psi = 0$. In our model, due to the introduction of the axion term, this corresponds to one of the two, real horizon solutions of $f(r) = 0$, which we call the axion-Reissner-Nordstr\"om (aRN) horizon, denoted by $r_{\text{aRN}}$. The other real solution is the typical event horizon, located at $r_{h}$ and for clarity, $r_{h} < r_{aRN}$ . Further details can be found in appendix \ref{s:rnv_appendix}. The significance of the ER bridge collapse can be summarised by Figure \ref{fig:psizero_psinonzero_gtt}. There we see that the $\psi = 0$, $g_{tt}$ function hits its zero precisely at $r = r_{\text{aRN}}$, while the full $\psi \neq 0$, $g_{tt}$ function does not, and exponentially decays after $r = r_{\text{aRN}}$ instead. This decay is what gives rise to the collapse nomenclature.

\begin{figure}[htb!]
\begin{center}
\includegraphics[width=8cm, height=6.5cm]{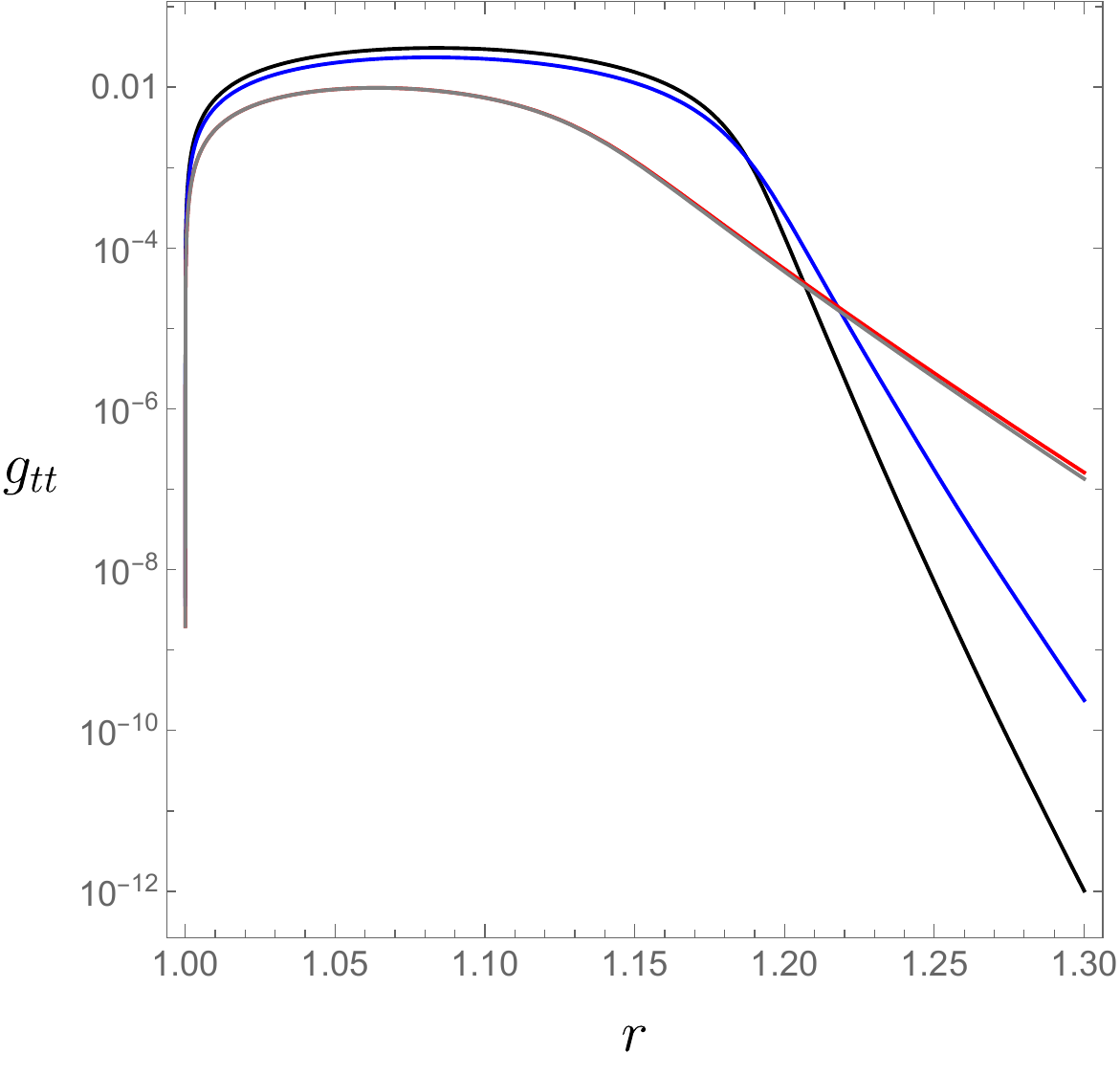}
\end{center}
\caption{ER collapse at various values of $\lambda$. The black, blue, red and grey curves respectively represent $\lambda = 0.01,1, 10$ and $1000$. Here $T/T_{c} \approx 0.9931$, $q =1$, $\gamma = 0$.}
\label{fig:er_collapse_gam0}
\end{figure}

For $\gamma = 0$ and  $\lambda \neq 0$, we identify the typical ER bridge collapse behaviour. Figure \ref{fig:er_collapse_gam0} demonstrates the collapse of the bridge under different $\lambda$ choices, all selected at approximately the same temperature $T/T_{c} \approx 0.9931$, with charge $q =1$. The collapse appears to become less severe at larger values of $\lambda$ and in fact tends to approach the same value (see the red and grey curves).

\begin{figure}
\centering
\begin{subfigure}{.5\textwidth}
	\centering 
	\includegraphics[width=1.0\linewidth]{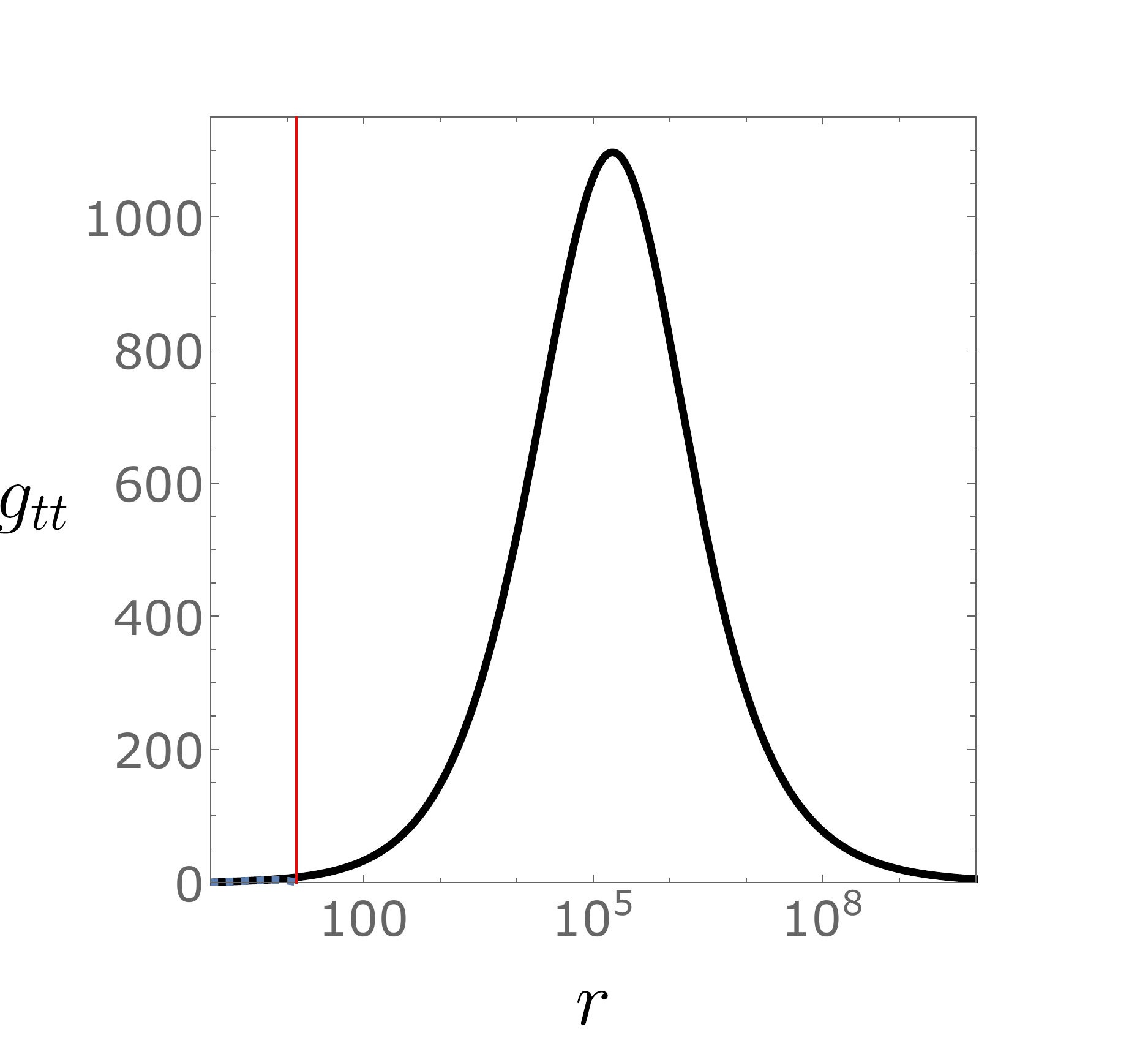}
\end{subfigure}%
\begin{subfigure}{.5\textwidth}
	\centering 
	\includegraphics[width=1.0\linewidth]{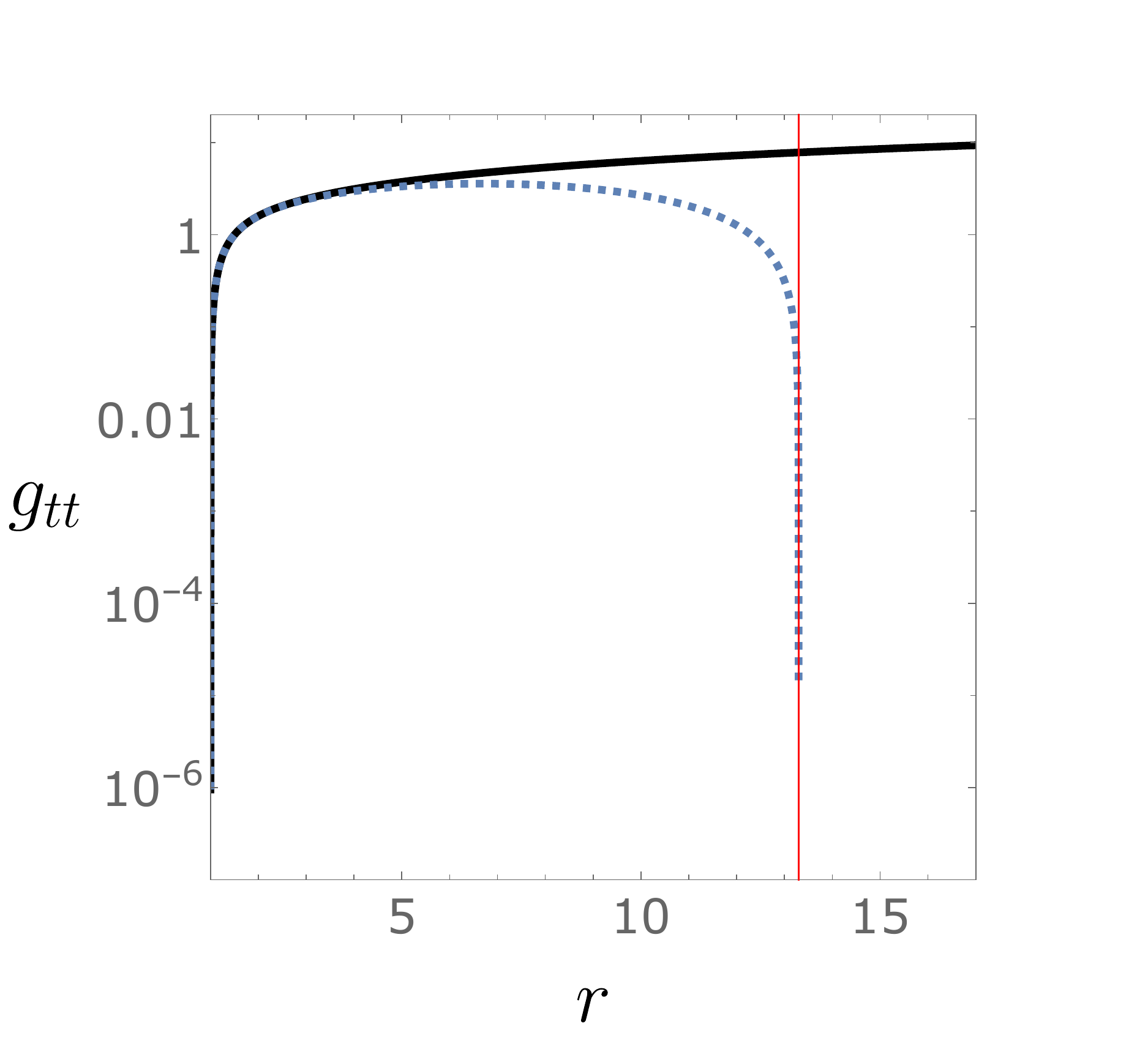}
\end{subfigure}
\caption{Plots of $g_{tt}$ for $\gamma = 30$. In the left figure we see how the full numerical $g_{tt}$ solution (black) grows from small $r$ all the way to $r \approx 10^{5}$ before it begins to collapse again. The right figure enhances the small $r$ region, showing the significant deviation of this $\gamma = 30$ solution from both the $\psi=0$ solution (dashed, blue) as well as the full $\gamma = 0$ numerical solutions seen in Figures \ref{fig:psizero_psinonzero_gtt} and \ref{fig:er_collapse_gam0}. Here $q=1$, $\lambda =0$ and $T/T_{c} = 0.9966$. }
\label{fig:er_collapse_lam1_gam_30}
\end{figure}

For $\lambda = 0$, $\gamma \neq 0$ we find seemingly different behaviour which delays the collapse of the ER bridge (and removes the Josephson oscillations, see section \ref{ssec:inv_tran_param_effects}). This comes about through an increase in $\gamma$ and as a concrete example we have taken $q=1$, $\lambda = 0$, $\gamma = 30$ and $T/T_{c} =0.9966$ in  Figure \ref{fig:er_collapse_lam1_gam_30}. Here we find that $g_{tt}$ undergoes positive growth over a large range of $r$ before once again decreasing. This contrasts with the typical sharp collapse around the aRN horizon\footnote{Note that in the numerical example provided, the aRN horizon is the standard inner horizon of Reissner-Nordstr\"om, since $\lambda = 0$.} demonstrated in Figures \ref{fig:psizero_psinonzero_gtt} and \ref{fig:er_collapse_gam0} for $\gamma = 0$. Analytic expressions describing the collapse of the ER bridge were first derived in \cite{Hartnoll:2020rwq} and later adapted to different models in \cite{Hartnoll:2020fhc} and \cite{Dias:2021afz}, where upon numerical inspection, those terms determined to be unimportant in the equations of motion were removed. The numerical plot in Figure \ref{fig:er_collapse_lam1_gam_30} indicates how terms that were previously negligible in the original collapse-to-Josephson oscillation epoch, become important for large $\gamma$, as we no longer see the sharp ER bridge collapse around the aRN horizon. In section \ref{ssec:inv_tran_param_effects}, we will plot the difference in these terms for alternative $\gamma$, and further discuss the geometry at large $\gamma$, showing that after large-$r$ collapse, it is well fitted to a Kasner regime.

\subsection{Emergence of Kasner geometry}
\label{ssec:kasner_geometry}

We begin by describing how numerical solutions led to the notion of flow from AdS to Kasner geometry in the holographic superconductor model. The Kasner flow argument first appeared in \cite{Frenkel:2020ysx} and was then applied to holographic superconductors in  \cite{Hartnoll:2020fhc} where numerical methods were used to solve the full set of equations of motion. From these numerical solutions, it was then inferred what terms in the equations were negligible in each epoch to achieve an analytic approximation of the fields. For the large-$r$ epoch, the analytic approximations of the metric functions were found to be\footnote{To be clear, these solutions are based on the original holographic superconductor model with $\lambda = 0, \gamma = 0$.}
\begin{align}
\label{eqn:BtH_kasner_function_form}
\begin{split}
f(r) = - f_{K} r^{3 + \beta^2} + \dots\,,& \quad \chi(r) = 2 \beta^2 \log{r} + \chi_{K} + \dots \,, \\
\psi(r) =  \sqrt{2} \beta \log{r} + \psi_{K}+ \dots \,,& \quad \phi(r) = \phi_{K_{1}}r^{1-\beta^2} + \phi_{K_{2}} + \dots 
\end{split}
\end{align}
where, $\psi_{K}$, $f_{K}$, $\chi_{K}$, $\phi_{K_{1}}$, $\phi_{K_{2}}$ and $\beta$ are constants. In this form the Kasner geometry becomes apparent with $\beta$ playing an important role. To see how, the radial coordinate is recast as $r = \tau^{-2/(3 + \beta^2)}$ and substituted into \eqref{eqn:TM_metric_ansatz} and \eqref{eqn:BtH_kasner_function_form}. The emergent metric is that of a Kasner universe \cite{Kasner:1921zz} given by
\begin{equation}
\label{eqn:BtH_kasner_metric}
ds^2 = - d\tau^2 + a_{t} \tau^{2 p_{t}}dt^2 + a_{x} \tau^{2 p_{x}}(dx^2 + dy^2) \,,
\end{equation}
with $a_{t}$ and $a_{x}$ constants. $p_{t}$ and $p_{x}$ are known as the Kasner exponents and adhere to the following conditions: $p_{t} + 2p_{x} = 1$ and $p_{\psi}^2 + 2 p_{x}^2 + p_{t}^2 = 1$. Upon making this transformation, the exponents are defined purely in terms of $\beta$
\begin{equation}
\label{eqn:BtH_kasner_exponents}
p_{x} = \frac{2}{3 + \beta^2}\,, \quad p_{t} = \frac{\beta^2 - 1}{3 + \beta^2}\,, \quad p_{\psi} = \frac{2\sqrt{2} \beta}{3 + \beta^2}\,.
\end{equation}
Here $p_{\psi}$ is defined through the redefinition of $\psi(r)$ in terms of $\tau$, i.e. $\psi(r) = \beta \sqrt{2} \log{(r)} \to \psi(\tau) = - p_{\psi} \log{\tau} + \psi_{\tau}$ with $\psi_{\tau}$ a constant.
This is therefore how the metric solution flows from an AdS geometry at the UV boundary to a Kasner geometry\footnote{This connection only holds true at radial distances for which \eqref{eqn:BtH_kasner_function_form} are valid.}. In addition to this discovery, there are further potential changes to the geometry known as Kasner inversions and transitions.

\subsection{Inversions, transitions and parameter effects}
\label{ssec:inv_tran_param_effects}

Kasner inversions can be readily identified from the full numerical solutions but can also be analytically identified \cite{Hartnoll:2020fhc} by the limiting behaviour the derivative of scalar field exhibits
\begin{equation}
\label{eqn:BtH_kas_asymp_psi_alpha_inversion_limit}
\frac{r \psi'}{\sqrt{2}} \to \frac{1}{\beta} \text{ as } r > r_{\text{inv}}, \quad \text{ and } \quad \frac{r \psi'}{\sqrt{2}} \to \beta \text{ as } r < r_{\text{inv}} \,,
\end{equation} 
where $\beta$ is as above and $r_{\text{inv}}$ is a constant and can be thought of as the ``radial position of inversion" i.e. where the inversion is localised. It is this limiting behaviour that gives rise to the inversion terminology. Put simply, if an inversion occurs, there are two different values for the Kasner exponents (and so two different Kasner geometries) corresponding to before and after said inversion.

\begin{figure}[htb!]
\begin{center}
\includegraphics[width=15.5cm, height=6cm]{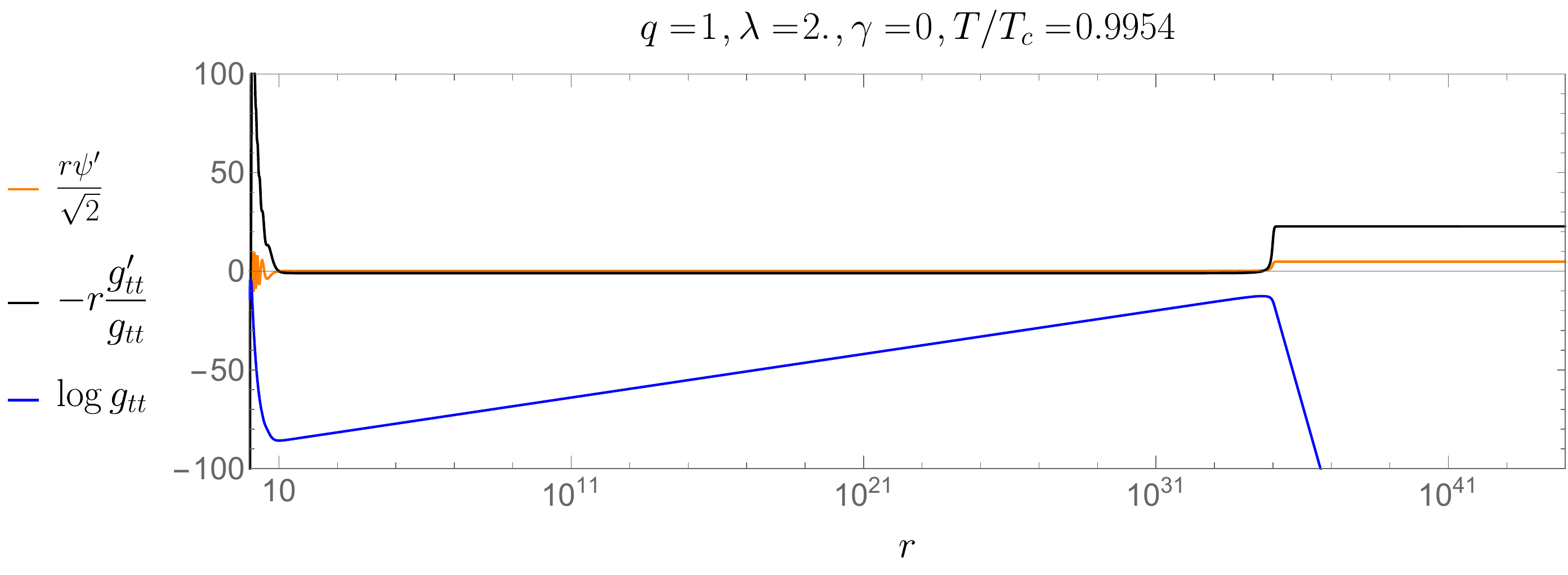}
\end{center}
\caption{A wide ranging plot of the interior ($r >1$) of the modified holographic superconductor, with $q=1, \lambda = 2, \gamma = 0$ and $T/T_{c} \approx 0.9954$. An inversion in this particular case is located at $r \approx 10^{35}$, which is clear from the three functions plotted: $\log{(g_{tt})}$ begins to collapse here once again, while $r \psi ' / \sqrt{2}$ and $-r g_{tt}'/g_{tt}$ become alternative constant values either side of this collapse point.}
\label{fig:lam_non_zero_inversion}
\end{figure}

\begin{figure}[htb!]
\begin{center}
\includegraphics[width=10cm, height=7cm]{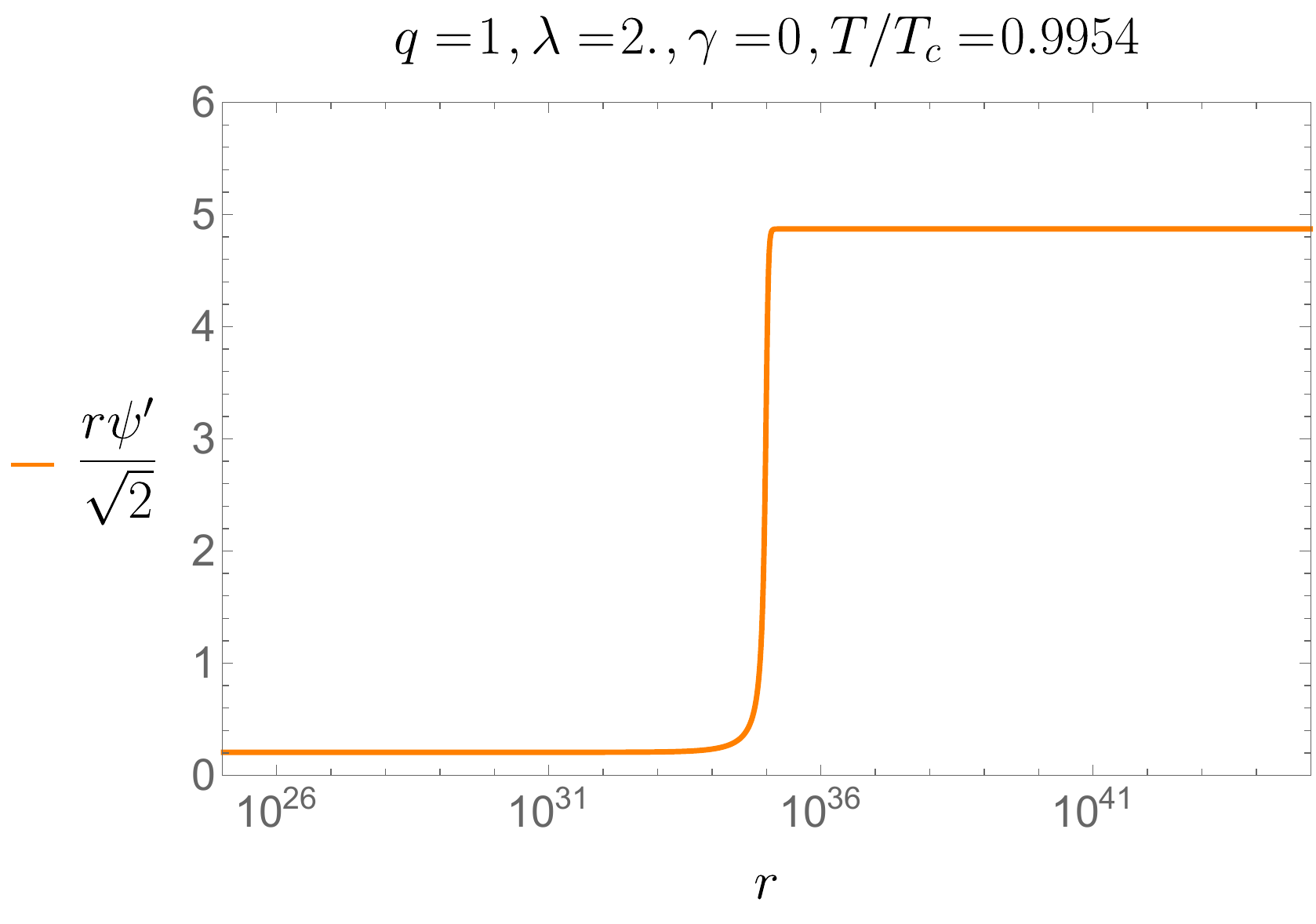}
\end{center}
\caption{An enhanced view of $r \psi'/ \sqrt{2}$, close to inversion. Before inversion $\beta_{\text{before}} = 0.20527901$, after inversion $\beta_{\text{after}} = 4.87141857$. The numerics therefore show good agreement to the inversion limit formula, with error $(1/\beta_{\text{before}} - \beta_{\text{after}}) \approx 9 \times 10^{-8}$. Once again, here  $q=1, \lambda = 2, \gamma = 0$ and $T/T_{c} \approx 0.9954$.}
\label{fig:lam_non_zero_inversion2}
\end{figure}

A natural question that arises for our model is, what are the effects of the new parameters and do the Kasner geometry and inversions remain? Starting with just the axion parameter $\lambda_{0}$ switched on, we argue that by the equations of motion evaluated on the approximate analytic solutions of \eqref{eqn:BtH_kasner_function_form}, the axion's contribution is negligible and so the geometry remains a Kasner universe at large-$r$. The relevant equation \eqref{eqn:TM_EOM4}, is given below for reference with $K(X) = X = \lambda_{0}^{2} r^2$ written explicitly 
\begin{equation}
\label{eqn:BtH_EOM4}
\frac{6}{r^2}-\frac{6}{r^2 f}-\frac{2 f'}{r f}+\frac{L^2 m^2 \psi^2}{r^2 f}+\frac{\lambda_{0} ^2}{f}+\frac{q^2 e^{\chi} \psi^2 \phi^2}{f^2}+ \frac{r^2 e^{\chi} \left(\gamma  \psi^2+1\right) (\phi ')^2}{2 f}+(\psi ')^2 = 0\,.
\end{equation}
Substituting \eqref{eqn:BtH_kasner_function_form} into the equation above and setting $\gamma = 0$, the left hand side expression reduces to 
\begin{align}
\label{eqn:BtH_reduced_EOM4_with_functions}
\begin{split}
\text{LHS} = c r^2 + \lambda_{0}^2 - \frac{6}{r^2} & +\frac{L^2 m^2 \left( \psi_{K} + \sqrt{2} \beta  \log{(r)} \right)^2}{r^2}  \\
 &- \frac{1}{f_{K}}e^{\chi_{K}} q^2 r^{-3 +\beta^2} \left( \phi_{K_{2}} +  r^{1-\beta^2} \phi_{K_{1}} \right)^2 \left(\psi_{K} + \sqrt{2} \beta \log{(r)} \right)^2 \,,
 \end{split}
\end{align}
where $c$ in the above equation is a constant, specifically $c = \frac{1}{2}e^{\chi_{K}}(\beta^2 -1)^2 \phi_{K_{1}}^2$. Since the $r$ dependence in the expression has powers governed by $\beta$, there is a range over which the dominating terms change. If $\beta^2 >5$, the term proportional to $r^{-3 + \beta^2}\log{(r)}^2$ will dominate at large-$r$, while the first term $c r^2 $ would be subleading and $\lambda_{0}^2$ would be negligible. On the other hand, taking for example $\beta^2 < 4$, the $c r^2$ term now dominates and the remaining terms become subleading. Therefore, within $ 4 < \beta^2  < 5$ there exists a dominance crossover. However, the salient point is that since the $\lambda_{0}$ term has no $r$ dependence, its contribution can be deemed negligible at large-$r$ either side of the $\beta$ crossover. To further validate this negligibility, a numerical example of Kasner inversion is provided in Figures \ref{fig:lam_non_zero_inversion} and \ref{fig:lam_non_zero_inversion2}. 

Figure \ref{fig:lam_non_zero_inversion} introduces an overview of the interior of our modified holographic superconductor model, from the event horizon at $r = 1$ towards the singularity/large-$r$. It demonstrates the three main epochs via three key functions: $\log{(g_{tt})}$, $ r \psi ' /\sqrt{2}$  and $-r g_{tt}'/ g_{tt}$, in blue, orange\footnote{This function is predominantly used for clarity, since it becomes constant in the Kasner epoch, before and after inversion.} and black respectively. The epoch between $r \approx 1$ and $r \approx 10$ contains the Josephson oscillations and the typical ER bridge collapse. Once the oscillations settle, an intermediate epoch is entered from $r \approx 10$ to $r \approx 10^{35}$. Then at $r \approx 10^{35}$ a secondary collapse occurs, after which the final epoch is entered. Figure \ref{fig:lam_non_zero_inversion2} enhances the region where the secondary collapse occurs and makes clear the constant behaviour of $r \psi ' / \sqrt{2}$ either side of inversion, as expressed in equation \eqref{eqn:BtH_kas_asymp_psi_alpha_inversion_limit}. Here both figures set\footnote{This selection of parameters has been chosen since it provides an easily visible inversion at large-$r$. } $q =1$, $\lambda = 2$, $\gamma = 0$ and $T/T_{c} \approx 0.9954$. If the axion term is indeed trivial in the Kasner epoch, then the inversion behaviour of the scalar field should remain valid with this choice of non-zero $\lambda$. To ensure we record the two limiting constant values of $r \psi' /\sqrt{2}$ accurately, the numerical solution is evaluated at $r = 10^{24}$ and $r = 10^{42}$ i.e. radial positions sufficiently before and after $r_{\text{inv}}$ but still within the large-$r$ regime, far from Josephson oscillations. We find the error to be of order $10^{-8}$ which supports the triviality of $\lambda$ in the Kasner epoch.

\begin{figure}[htb!]
\begin{center}
\includegraphics[width=15.5cm, height=6cm]{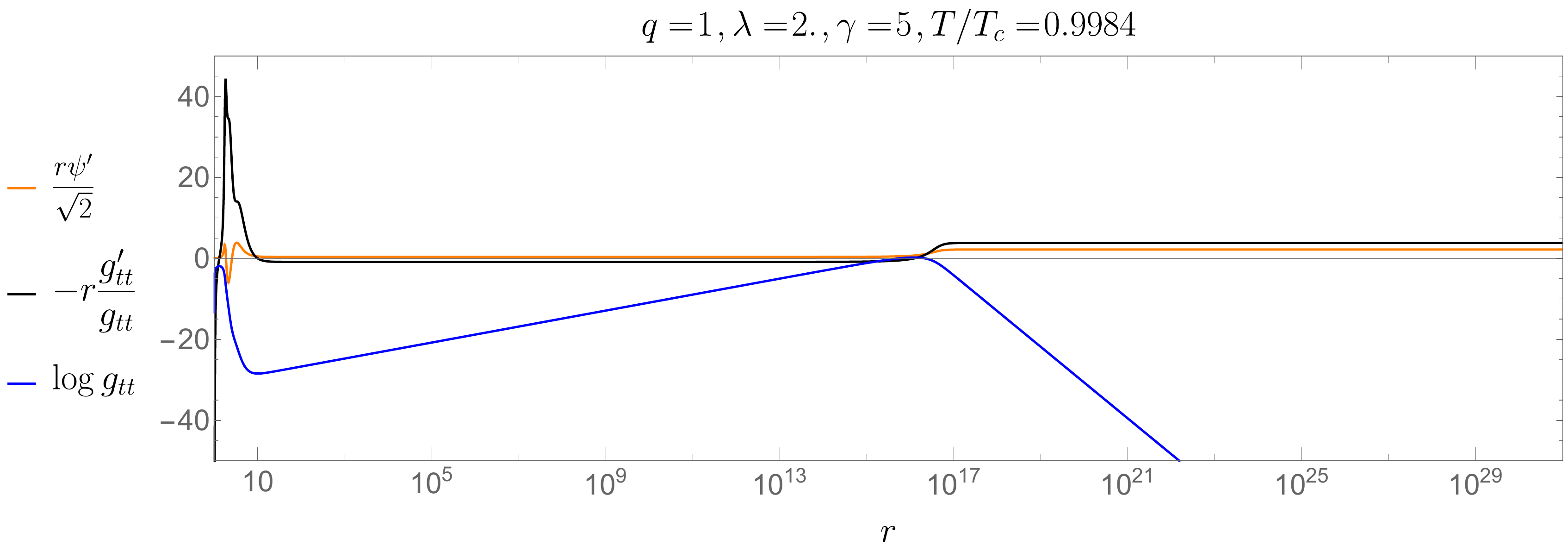}
\end{center}
\caption{The plot shows the three functions of interest for $q=1$, $\lambda = 2$, $\gamma = 5$ and $T/T_{c} = 0.9984$. Around $r \approx 10^{16}$ there is another collapse of the ER bridge, indicated in blue, which brings about a Kasner transition.}
\label{fig:gam_lam_non_zero_inversion}
\end{figure}

\begin{figure}[htb!]
\begin{center}
\includegraphics[width=11cm, height=7.5cm]{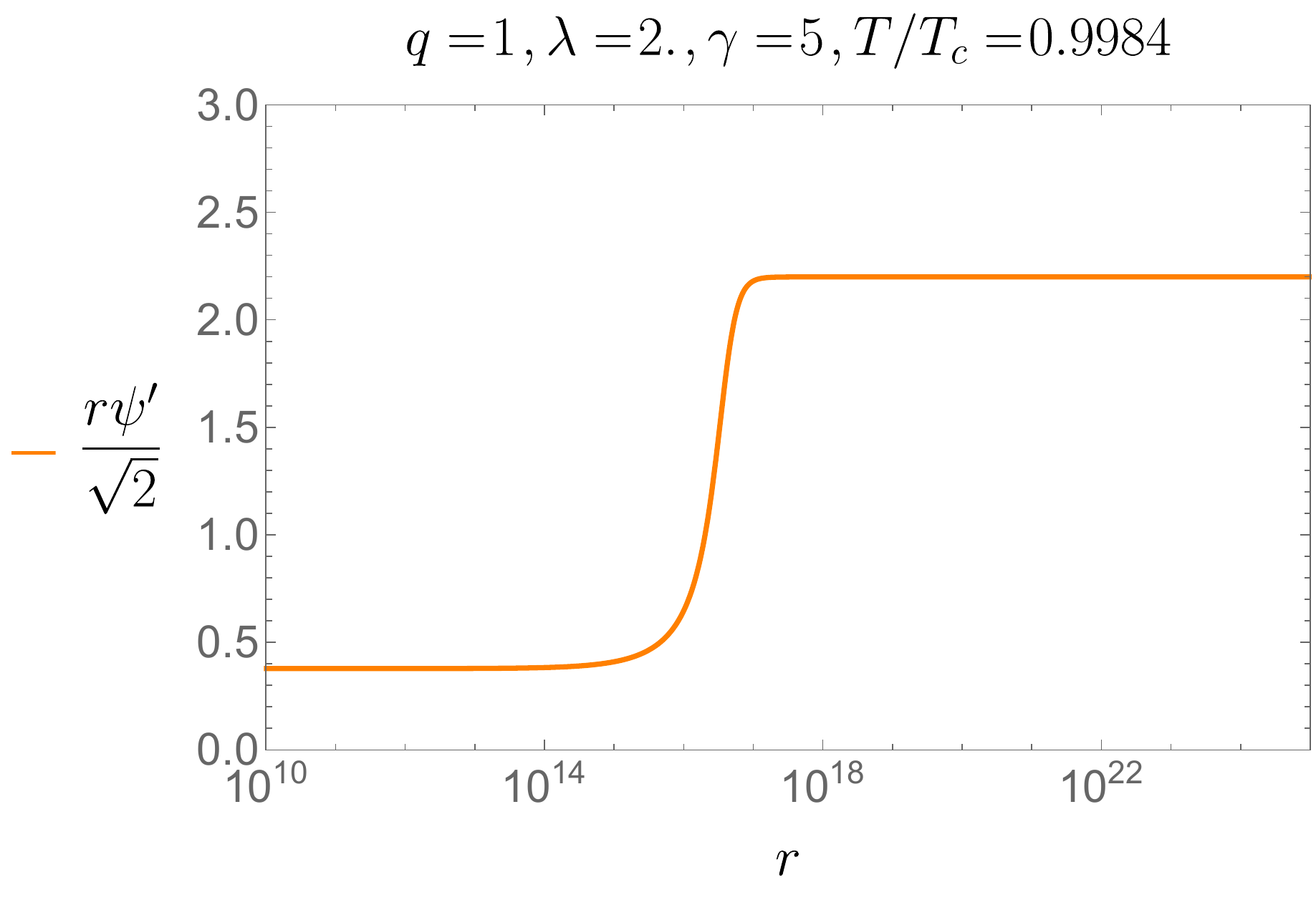}
\end{center}
\caption{An enhanced plot of $r \psi'/\sqrt{2}$ close to the transition for the same values $q=1$, $\lambda =2$, $\gamma = 5$ and $T/T_{c} = 0.9984$. Before transition $r \psi'/\sqrt{2} = 0.3778567$, after transition $r \psi'/ \sqrt{2} = 2.1993928$, which indicate this is not an inversion. Again, here $q = 1$, $\lambda = 2$, $\gamma = 5$ and $T/T_{c} = 0.9984$.}
\label{fig:gam_lam_non_zero_inversion2}
\end{figure}

Taking both $\lambda$ and $\gamma$ non-zero, we can identify \textit{Kasner transitions}, a feature which again occurs when the ER bridge undergoes a secondary collapse. While similar, these are not the same as Kasner inversions. The reason being that the EMS term complicates the equation that describes $\psi$ and as such, the simple inversion behaviour of \eqref{eqn:BtH_kas_asymp_psi_alpha_inversion_limit} no longer exists. Interestingly, \cite{Dias:2021afz} explored the inside of asymptotically flat black holes, by introducing an EMS term and it was here transitions were first found. Under the same methodology, they showed that the analytic equations are the same as \cite{Hartnoll:2020fhc} but with additional contributions from this EMS term. Since $\lambda$ is negligible (along with the other terms found in both works such as the cosmological constant/2d spherical curvature) then our model directly reduces to that of \cite{Dias:2021afz}, in the large-$r$ Kasner epoch. 
An example of a transition is provided in Figures \ref{fig:gam_lam_non_zero_inversion} and \ref{fig:gam_lam_non_zero_inversion2} generated for $q=1$, $\lambda =2$, $\gamma = 5$ and $T/T_{c} \approx 0.9984$. The general behaviour is clear in Figure \ref{fig:gam_lam_non_zero_inversion}, beginning with ER bridge collapse, followed by Josephson oscillations that pass into a Kasner regime. Eventually, around $r \approx 10^{16}$ the Kasner transition occurs. Figure \ref{fig:gam_lam_non_zero_inversion2} enhances the point of transition and from the numerical error, shows that it cannot be a simple inversion, as occurs when $\gamma = 0$.

\begin{figure}[htb!]
\begin{center}
\includegraphics[width=15.5cm, height=6cm]{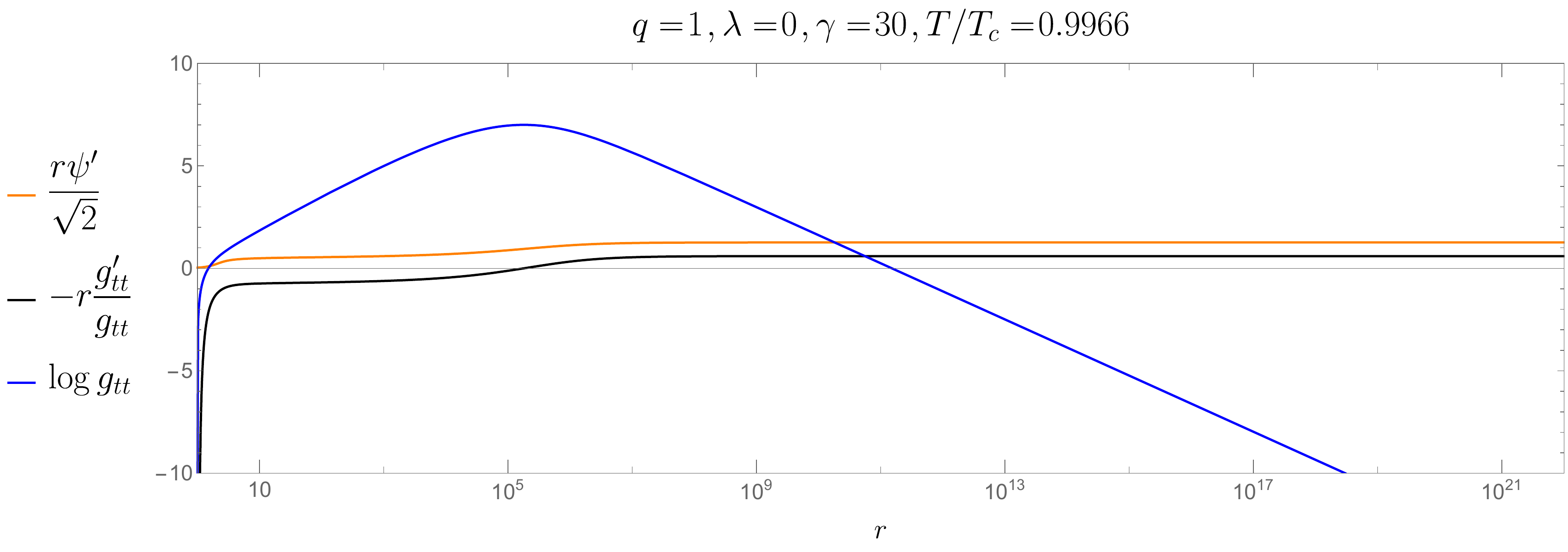}
\end{center}
\caption{A behind-the-horizon/interior plot, displaying functions $\log{(g_{tt})}$, $r \psi'/ \sqrt{2}$ and $- r g_{tt}' /g_{tt}$, with $q=1$, $\lambda=0$, $\gamma = 30$ and $T/T_{c} =0.9966$. It demonstrates an initial growth in $\log{(g_{tt})}$ that extends far beyond the aRN horizon. Its maximum point at $r \approx 10^5$ and subsequent collapse, is also clear by the change in the other two functions. Noticeably, the $r \psi'/\sqrt{2}$ orange curve at small-$r$ shows the absence of Josephson oscillations.}
\label{fig:er_bridge_growth}
\end{figure}

We now return to the analysis of large $\gamma$ to explain the observations made in section \ref{ssec:er_bridge} and Figure \ref{fig:er_collapse_lam1_gam_30}, and what this means for the geometry. Figure \ref{fig:er_bridge_growth} displays the same numerical solution as Figure \ref{fig:er_collapse_lam1_gam_30}, but now depicting functions $\log{(g_{tt})}$, $ r \psi ' / \sqrt{2}$ and $-r g_{tt}'/ g_{tt}$, over a wide range of $r$.
The functions show two clear features. The first is the absence of both the Josephson oscillations and collapse around the aRN horizon. The second is the repositioning of the  collapse of $g_{tt}$ (here $\log{(g_{tt})}$) to large-$r$, which leaves its effect imprinted on $ r \psi '/\sqrt{2}$ and $-r g_{tt}'/ g_{tt}$.

\begin{figure}[htb!]
\begin{center}
\includegraphics[width=16cm, height=5cm]{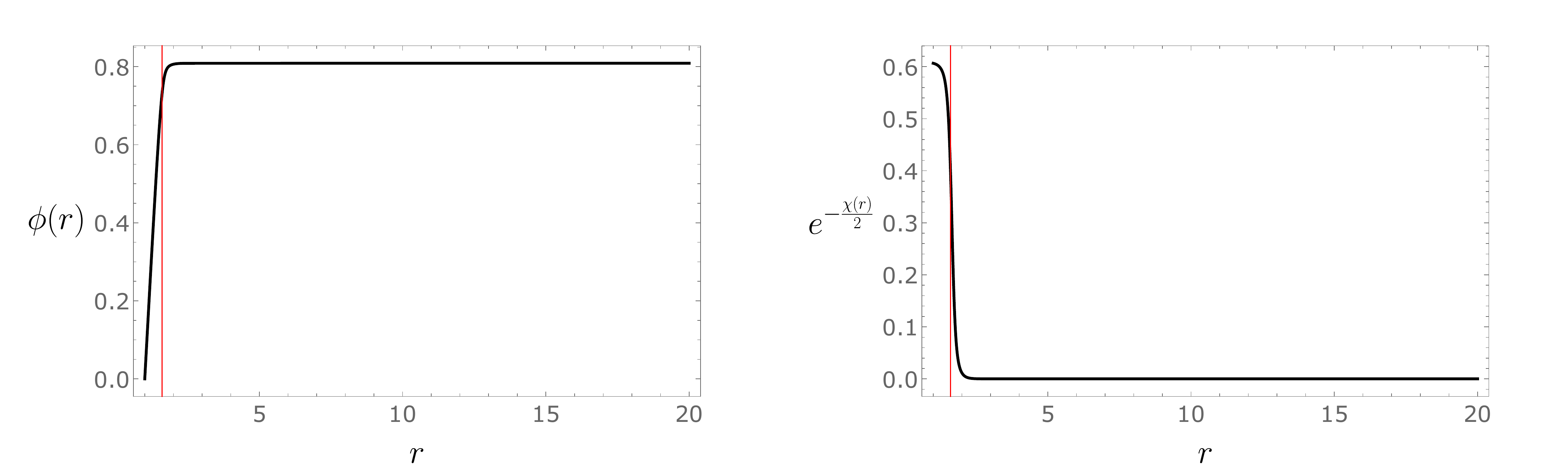}
\end{center}
\caption{Functions $\phi(r)$ and $e^{- \frac{\chi}{2}}$ for $\gamma = 1$. The red line is the location of the aRN horizon. Here we see that $\phi(r)$ becomes effectively constant after the aRN horizon, while  $e^{- \frac{\chi}{2}}$ drops off to zero. Here, $\lambda = 0$, $q=1$ and $T/T_{c} \approx 0.9966$.}
\label{fig:gamma1functions}
\end{figure}

\begin{figure}[htb!]
\begin{center}
\includegraphics[width=16cm, height=5cm]{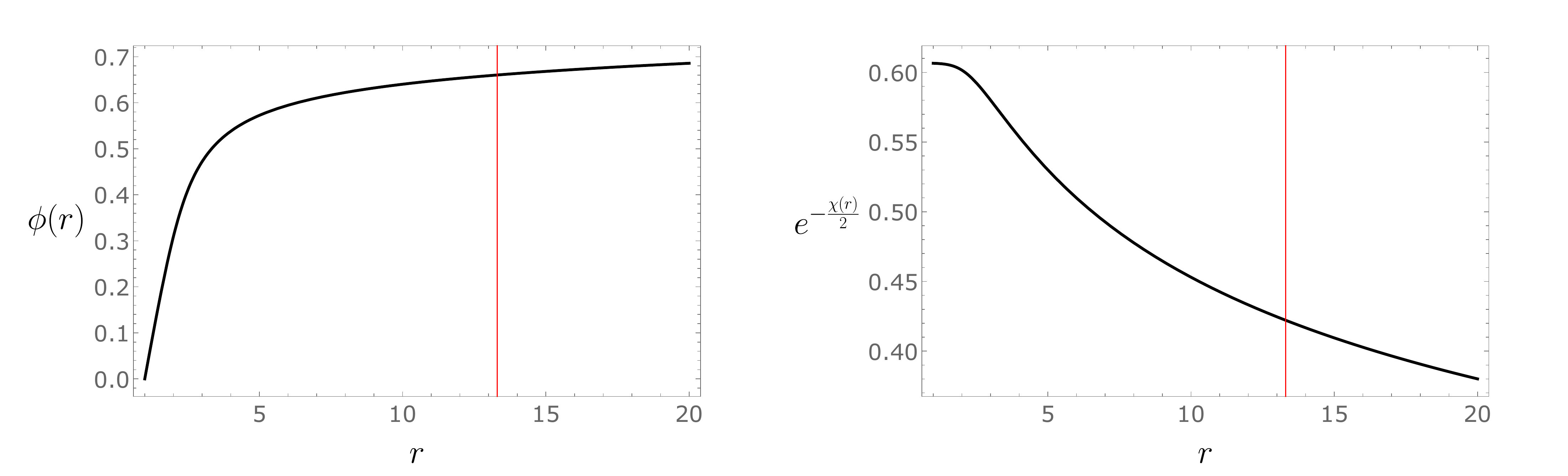}
\end{center}
\caption{Functions $\phi(r)$ and $e^{- \frac{\chi}{2}}$ for $\gamma = 30$. The red line is the location of the aRN horizon. Here we see that $\phi(r)$ is no longer constant soon after the aRN horizon, but rather is a function of $r$. As for, $e^{- \frac{\chi}{2}}$ this does not drop to zero soon after the aRN horizon. As a result, terms involving these functions in the equations of motion can no longer be neglected. Here, $\lambda = 0$, $q=1$ and $T/T_{c} \approx 0.9966$.}
\label{fig:gamma30functions}
\end{figure}

The first feature appears because the terms in the equations of motion containing $\gamma$ are no longer negligible at such radial distances. Checking the numerical solution of $\phi$ and $e^{-\chi/2}$ at large $\gamma$ shows that $\phi$ is no longer constant, immediately after the aRN horizon but rather it grows with $r$, while $e^{- \chi/2}$ no longer decays quickly to zero after the aRN horizon. These two criteria are crucial to the appearance (or lack thereof) of Josephson oscillations, as is pointed out by the analytic approximations found in both \cite{Hartnoll:2020fhc} and \cite{Dias:2021afz} that describe them. To compare the behaviour of these functions when $\gamma$ is small to when $\gamma$ is large, Figure \ref{fig:gamma1functions} shows that for $\gamma = 1$ we have that $\phi(r) \approx \text{const.}$ and $e^{-\chi/2} \approx 0$ after the aRN horizon. This is typical of the standard Josephson oscillation epoch, where upon using these approximations, a Bessel function form of the scalar field $\psi(r)$ can be obtained. On the other hand, Figure \ref{fig:gamma30functions} shows that for $\gamma = 30$, the behaviour around the aRN horizon is far different. As mentioned, $\phi(r)$ is no longer constant but scales with $r$, while $e^{-\chi/2}$ is no longer well approximated as being zero either. In both of these figures, $\lambda = 0$, $q=1$ and $T/T_{c} \approx 0.9966$.

The second feature is much like the previous inversion and transition behaviour seen prior, since there is a change in the $r \psi' / \sqrt{2}$ and $-r g_{tt}' /g_{tt}$ functions once $\log{(g_{tt})}$ starts to decrease at $r \approx 10^5$. However, the geometry is only well modelled as a Kasner universe after the collapse point and not before. Therefore, this point is neither an inversion or transition in the previously defined sense, but it still marks the change to a Kasner geometry. To validate this large-$r$ Kasner behaviour, the analytic Kasner expressions are adapted to include the additional modifications due to $\gamma$ (see Appendix \ref{s:simp_eqn_appendix} for the resulting equations of motion). The corrected form, as explained in \cite{Dias:2021afz}, amounts to a change in the gauge field function
\begin{equation}
\label{eqn:Bth_dhs_gauge_function}
\phi(r)  = \phi_{K_{1}} +  \frac{\phi_{K_{2}} r^{1- \beta^2}}{1 + 2 \gamma \beta^2 \log^2{r}}  \,.
\end{equation}

\begin{figure}[htb!]
\begin{center}
\includegraphics[width=15cm, height=10cm]{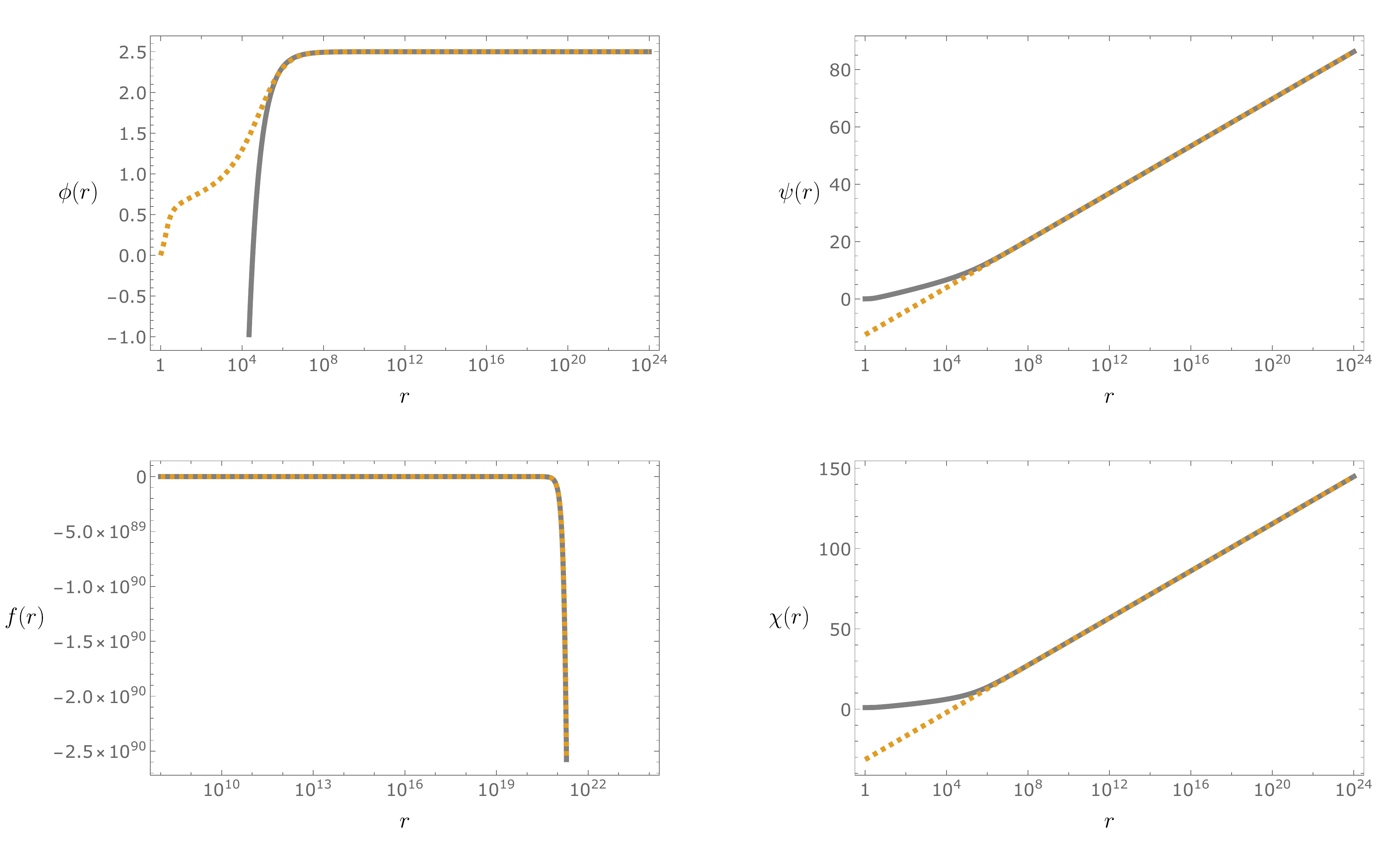}
\end{center}
\caption{The four fields $\phi$, $\psi$, $f$ and $\chi$ under consideration in the Kasner epoch. The grey line refers to the full numerical solution, while the orange dashed line is the analytic expression from the simplified equations of motion. There is good agreement between the numerics and analytic approximations for all functions, when evaluated sufficiently past the point of collapse, $r \approx 10^5$. Here  $q=1$, $\lambda=0$, $\gamma = 30$ and $T/T_{c} =0.9966$.}
\label{fig:four_kas_functions}
\end{figure}

Fitting the functions $f$, $\chi$ and $\psi$ from equation \eqref{eqn:BtH_kasner_function_form} and $\phi$ from \eqref{eqn:Bth_dhs_gauge_function} above, to the full numerical solutions shows good agreement past the large-$r$ collapse point. This is displayed in Figure \ref{fig:four_kas_functions} where the grey curve is the full numeric solution and the dashed orange curve is the analytic function. Noticeably, the fit is not a good approximation pre-collapse (before $r \approx 10^5$) marking a significant change compared to the small $\gamma$ solutions, where we would either observe oscillations or another Kasner universe.

To summarise, for large $\gamma$, the typical ER bridge collapse and Josephson oscillation behaviour is no longer exhibited at and around the aRN horizon. This can be seen numerically but also argued from the analytic standpoint. The reason being is that when $\gamma$ is large, $\phi$ and $e^{-\chi/2}$ can no longer be dropped from the equations of motion, and hence $\psi$ is not approximated by a Bessel function. Moving outward, from the aRN horizon to the new point of large-$r$ collapse, the form of the solution is simple but not Kasner, while past the point of collapse it becomes Kasner as demonstrated in Figure \ref{fig:four_kas_functions}. 

\section{Results for \texorpdfstring{$K(X) = X^3$}{TEXT}}
\label{sec:X_cubed_results}

Potentials $K(X) = X^n$ with $n< 5/2$, such as the $n=1$ model discussed in the previous sections \ref{sec:phase_diagrams} and \ref{sec:behind_the_horizon}, correspond to explicit translational symmetry breaking. This is because the leading term in the axion's near-boundary expansion is spatially dependent (equivalent to saying that there is an $x^{I}$-dependent source). On the other hand, when $n> 5/2$, the translational symmetry is spontaneously broken, with the axion becoming a spatial-coordinate dependent expectation value, see \cite{Baggioli:2021xuv,Alberte:2017oqx,Li:2018vrz} for further details. It is therefore interesting to explore the outcomes of the $K(X) = X^3$ model since it distinguishes itself from $K(X) = X$ in this way. The analysis in this section consists of producing the $X^3$ analogue of the $X$ model's condensate vs. temperature plots of Figure \ref{fig:phase_diagram_lambda}, as well as observing the effects of this higher power of $X$ on the metric functions behind the horizon, akin to the plots in Figure \ref{fig:lam_non_zero_inversion} and \ref{fig:lam_non_zero_inversion2}.

\begin{figure}
\centering
\begin{subfigure}{1.0\textwidth}
	\centering 
	\includegraphics[width=13cm, height=7.2cm]{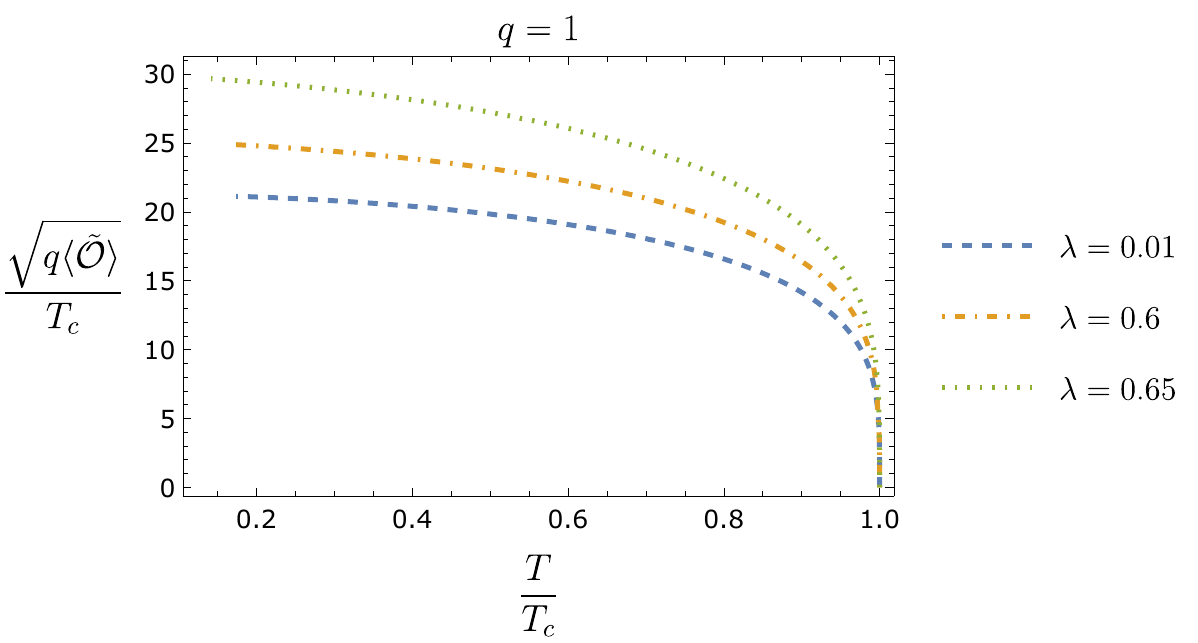}
	\vspace{0.3cm}
\end{subfigure}%

\begin{subfigure}{1.0\textwidth}
	\centering 
	\includegraphics[width=13cm, height=7.2cm]{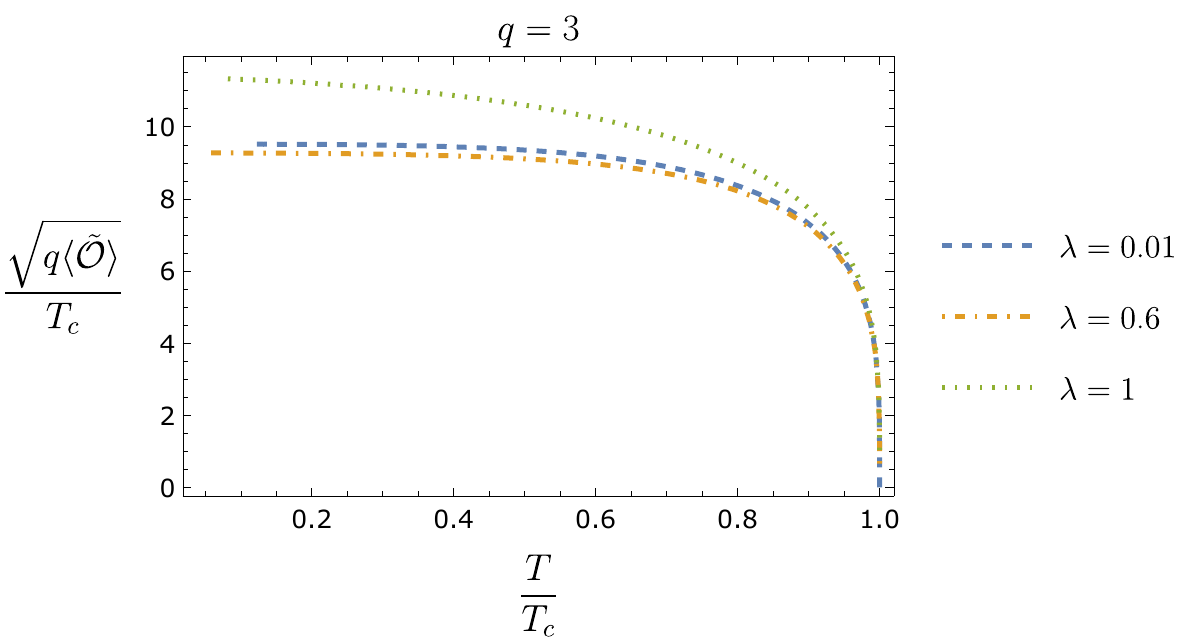}
	\vspace{0.3cm}
\end{subfigure}

\begin{subfigure}{1.0\textwidth}
	\centering 
	\includegraphics[width=13cm, height=7.2cm]{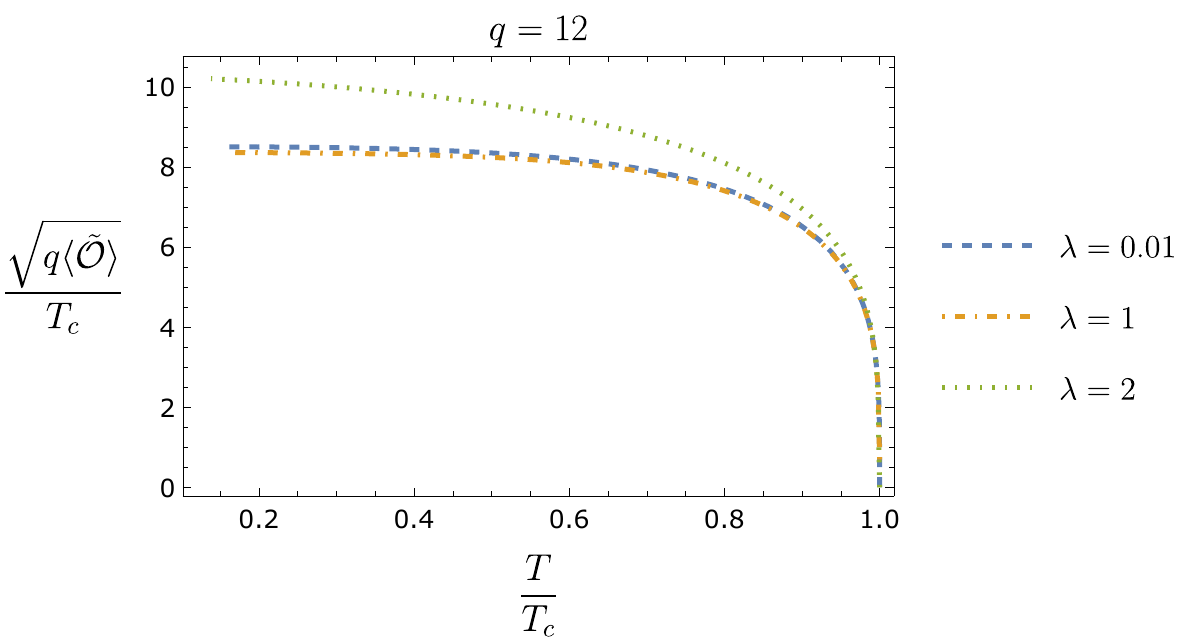}
\end{subfigure}
\caption{$K(X) = X^3$ model condensate plots for $q=1, 3$ and $12$ at $\gamma = 0$ and various $\lambda$. Much like the $K(X) = X$ case, increasing the charge leads to an overall decrease in the condensate towards the probe limit value. The $q=3, $ and $12$ plots additionally show that the condensate can decrease with an increase in $\lambda$. This only seems to appear at relatively small $\lambda$ while for large $\lambda$, the condensate generally increases. }
\label{fig:x3_phase_diagram_lambda}
\end{figure}

The $K(X) = X^3$ condensate plots are given in Figure \ref{fig:x3_phase_diagram_lambda} and  follow similar structure to those of the $K(X) = X$ case, in the sense that increasing the charge $q$ has the effect of reducing the condensate for all $\lambda$. The $q=3$ and $q=12$ plots for the $X^3$ model do appear to show new behaviour however, in that an initial reduction in the condensate can occur when small values of $\lambda$ are increased (see the $\lambda = 0.01$ vs. $\lambda = 0.6$ curves in the $q=3$ plot for example). For $\lambda$ greater than these small values, the condensate then appears to increase as seen in the previous model.

\begin{figure}[htb!]
\begin{center}
\includegraphics[width=15.5cm, height=6cm]{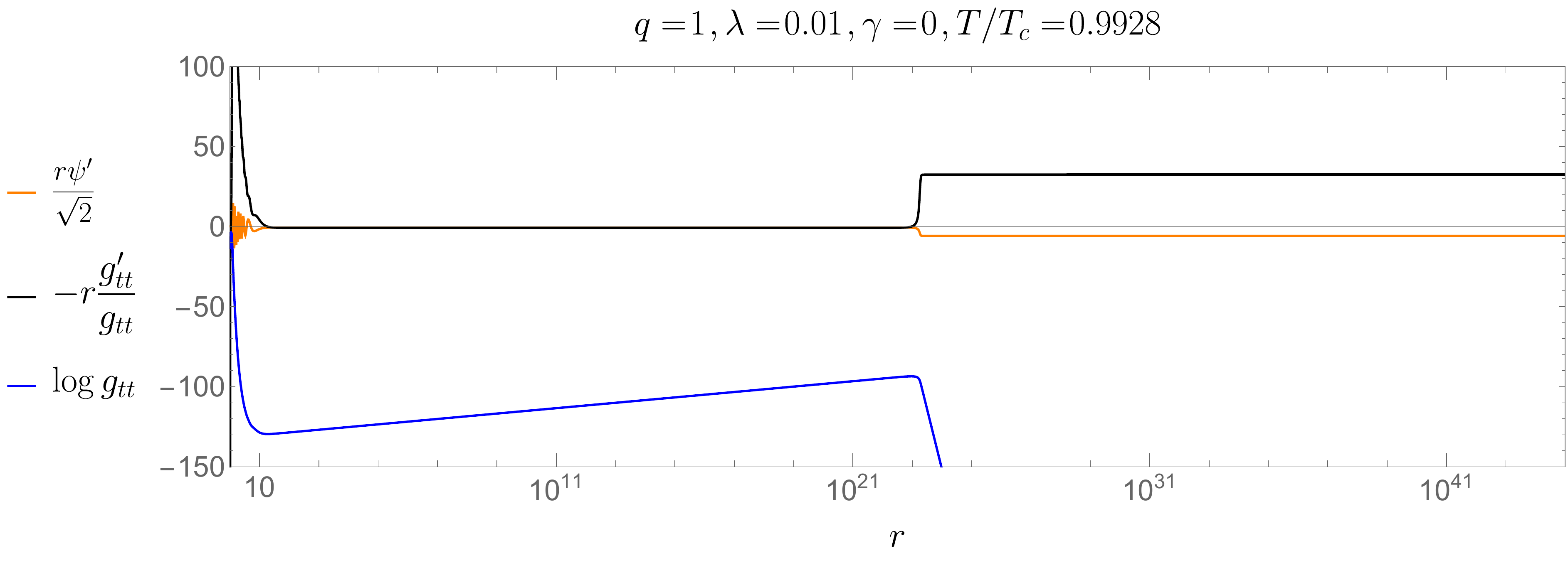}
\end{center}
\caption{Displays the interior of the $K(X) = X^3$ model, and presents functions $\log{(g_{tt})}$, $r \psi'/ \sqrt{2}$ and $- r g_{tt}' /g_{tt}$, with $q=1$, $\lambda=0.01$, $\gamma = 0$ and $T/T_{c} =0.9928$. The features observed in the $K(X) = X$ model still remain, with early Josephson oscillations and ER bridge collapse. However, the secondary collapse at $r \approx 10^{23}$ is not an inversion but a transition.}
\label{fig:x3behindhorizon}
\end{figure}

\begin{figure}[htb!]
\begin{center}
\includegraphics[width=11cm, height=7.5cm]{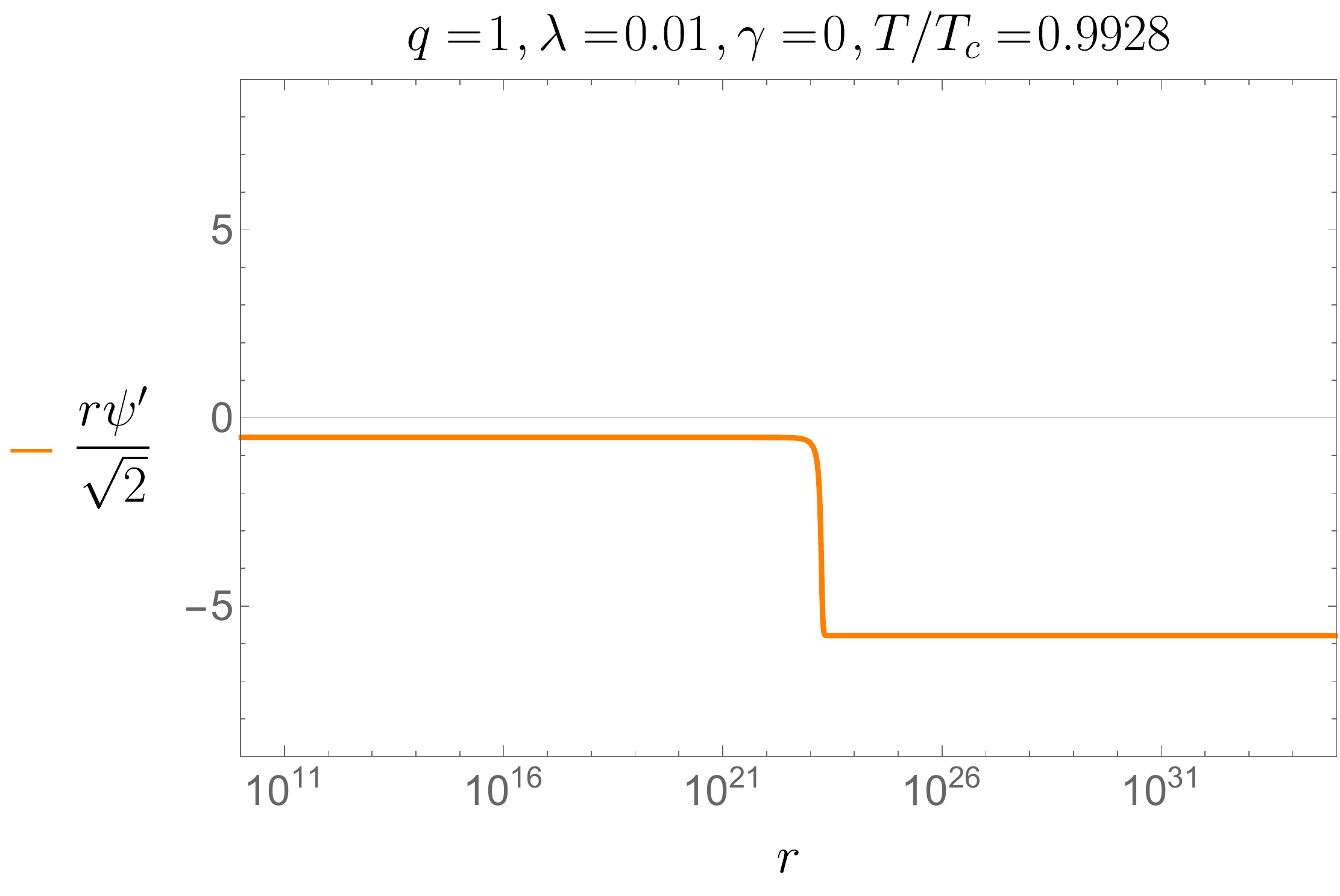}
\end{center}
\caption{Enhances the $K(X) = X^3$ model's  $r \psi'/ \sqrt{2}$ function at the point of secondary collapse. Here $q=1$, $\lambda=0.01$, $\gamma = 0$ and $T/T_{c} =0.9928$. Before transition $\beta_{\text{before}} = -0.518461$, after transition $\beta_{\text{after}} = -5.786356$, clearly indicating the simple limiting behaviour seen in the $K(X)=X$ model is not present in the $K(X) = X^3$ model.}
\label{fig:x3behindhorizonzoom}
\end{figure}

As for behind the horizon,  Figure \ref{fig:x3behindhorizon} demonstrates the typical functions for the $K(X) = X^3$ model, with $q =1, \lambda = 0.01, \gamma = 0$ and $T/T_{c} = 0.9928$. At small $r$, the Josephson oscillations and Einstein-Rosen bridge collapse can be seen, while a secondary collapse at $r \approx 10^{23}$ is also present. An enhanced plot of  $r \psi' / \sqrt{2}$ around this secondary collapse is provided in Figure \ref{fig:x3behindhorizonzoom} and upon checking its value either side, we find that unlike the $K(X) = X$ model, this behaviour is not a Kasner inversion but rather a Kasner transition. This shows that the simple limiting behaviour of \eqref{eqn:BtH_kas_asymp_psi_alpha_inversion_limit} does not hold for the $K(X) = X^3$ model, and is likely due to the $\lambda_{0}$ term in \eqref{eqn:BtH_reduced_EOM4_with_functions} gaining positive power of $r$ dependence and so it can no longer be deemed negligible at large-$r$.

\clearpage


\section{Discussion}
\label{s:discussion}
This paper has looked at how a holographic superconductor, modified by an axion and an EMS term, can affect the condensate and cause changes in the interior Kasner geometries. The condensate and its dependence on the parameters were first studied. The general results showing that as the axion parameter increases the condensate increases, as the EMS parameter increases the condensate decreases and as the charge increases, the value of the condensate progresses closer to its probe limit value. The interior of the black hole was then analysed with attention given to collapse and emergent Kasner properties. When the axion term is present for the $K(X) = X$ model, the large-$r$ geometry was argued to remain Kasner analytically and numerically, signifying the triviality of the axion behind the horizon at large-$r$. The Kasner geometry also remained for the $K(X) = X^3$ model but here the simple inversion behaviour after collapse no longer holds. Varying the EMS parameter saw somewhat different phenomena emerge. The Josephson oscillations are removed and the typical ER bridge collapse at the aRN horizon is shifted outward to a large radial value. This large-$r$ collapse point divides the geometry; non-Kasner before collapse and Kasner afterwards. 

To close out the discussion, a selection of future research directions that could build on our findings are mentioned. Having seen that large $\gamma$ solutions are non-Kasner before the large-$r$ collapse, via the numerics of Figure \ref{fig:four_kas_functions}, it would be interesting to arrive at analytic expressions for the geometry in this region. In a similar vein, a full analytic understanding of the absence of the oscillations and the accompanying description in terms of temperature would be desirable. A ``no-inner horizon" proof for this axion-EMS model seems feasible and could potentially be obtained by applying the familiar conserved quantity arguments found throughout the literature, for example \cite{Cai:2020wrp,Gubser:2009cg}. Exploring the physical reasoning for the change in condensate upon varying the axion and EMS parameters would be another natural step forward. It would be especially interesting to understand if there exists a specific limit or opportune scaling that could relate the condensate curves of different parameter values. Lastly, there exist valid alternatives to the selection of scalar field mass and so this could be explored as an additional parameter.

\vspace{0.5in}   \centerline{\bf{Acknowledgements}} \vspace{0.2in}

We thank the authors of \cite{Frenkel:2020ysx} for sharing their Mathematica notebook. We also thank the referees for their insightful comments. LS is supported by an STFC studentship. DV is supported by the STFC Ernest Rutherford grant ST/P004334/1.

\newpage
\appendix

\numberwithin{equation}{section}

\setcounter{equation}{0}

\section{aRN geometry at  \texorpdfstring{$\psi(r) = 0$}{TEXT}}
\label{s:rnv_appendix}
When $\psi(r) = 0$ the condensate disappears and the system enters normal phase. As such the consequent equations of motion become
\begin{subequations}
\begin{align}
\frac{1}{2}\phi' \chi' + \phi'' &= 0 \,, \label{eqn:App_normal_phase_EOM1} \\
\chi' &= 0 \,, \label{eqn:App_normal_phase_EOM2} \\
\frac{6}{r^2} - \frac{6}{r^2 f} + \frac{\lambda_{0}^2}{f} - \frac{2 f'}{f} + \frac{e^{\chi}r^2 (\phi')^2 }{2 f} &= 0 \,, \label{eqn:App_normal_phase_EOM3} 
\end{align}
\end{subequations}
which have solutions
\begin{align}
\label{eqn:App_black_factor}
f(r) &= 1 - \frac{r^3}{r_{h}^3} - \frac{r^2 \lambda_{0}^2}{2} + \frac{r^3 \lambda_{0}^2}{2 r_{h}}+ \frac{r^4 \rho^2}{4} - \frac{r^3 r_{h} \rho^2}{4} \,, \\
\chi(r) &= \text{const.} \,, \\
\phi(r) &= \mu  + r \rho \,,
\end{align}
where $\rho$ is the charge density, $\mu$ is the chemical potential. Upon setting $f(r) = 0$ one finds four roots, two complex and two real. The two real solutions are the event horizon $ r_{h}$ and the axion-Reissner-Nordstr\"om horizon $r_{\text{aRN}}$.

\section{Simplified equations of motion for Kasner epoch}
\label{s:simp_eqn_appendix}
Here the necessary simplifications that result in the $\gamma \neq 0$, $K(X) = X$ analytic Kasner functions are listed. The equations of motion  \eqref{eqn:TM_EOM1}-\eqref{eqn:TM_EOM4} are first rewritten as
\begin{subequations}
\begin{align}
r^2 e^{-\frac{\chi}{2}}\left[(1 + \gamma \psi^2)e^{\frac{\chi}{2}}\phi' \right]' - \frac{2 q^2 \psi^2 \phi }{f} &= 0 \,, \label{eqn:App_EOM1_alt} 
\\
r^2 e^{\frac{\chi}{2}}\left[ \frac{e^{-\frac{\chi}{2}}f \psi'}{r^2} \right]' + \left(-\frac{L^2 m^2}{r^2} + \frac{e^{\chi} q^2 \phi^2}{f} + \frac{e^{\chi} r^2 \gamma (\phi')^2}{2} \right)\psi  &=0 \,, \label{eqn:App_EOM2_alt}
\\
\frac{q^2  e^{\chi } \phi^2 \psi^2}{f^2} - \frac{\chi' }{r} +  (\psi ')^2 &= 0\,, \label{eqn:App_EOM3_alt}
\\
e^{\frac{\chi}{2}} r^4 \left[\frac{e^{-\frac{\chi}{2}} f}{r^3} \right]' + 3 - \frac{m^2 L^2 \psi^2}{2} - \frac{\lambda_{0}^2 r^2}{2} - \frac{1}{4}(1 + \gamma \psi^2) e^{\chi} r^4 (\phi')^2  &= 0\,. \label{eqn:App_EOM4_alt}
\end{align}
\end{subequations}
In the Kasner regime, the following terms are negligible and can be dropped: mass, charge, axion coupling, curvature. This leaves the following set of equations
\begin{subequations}
\begin{align}
r^2 e^{-\frac{\chi}{2}}\left[(1 + \gamma \psi^2)e^{\frac{\chi}{2}}\phi' \right]' = 0 \implies \phi' &= \frac{E_{0}e^{-\frac{\chi}{2}}}{1+ \gamma \psi^2} \,, \label{eqn:App_EOM1_simp} 
\\
r^2 e^{\frac{\chi}{2}}\left[ \frac{e^{-\frac{\chi}{2}}f \psi'}{r^2} \right]' + \left(\frac{e^{\chi} r^2 \gamma (\phi')^2}{2} \right)\psi  &=0 \,, \label{eqn:App_EOM2_simp}
\\
- \frac{\chi' }{r} +  (\psi ')^2 &= 0\,, \label{eqn:App_EOM3_simp}
\\
e^{\frac{\chi}{2}} r^4 \left[\frac{e^{-\frac{\chi}{2}} f}{r^3} \right]'  - \frac{1}{4}(1 + \gamma \psi^2) e^{\chi} r^4 (\phi')^2  &= 0\,. \label{eqn:App_EOM4_simp}
\end{align}
\end{subequations} 
Note that as discussed in section \ref{ssec:inv_tran_param_effects}, $e^{-\chi/2}$ being small can lead to alternative simplifications that would result in a Bessel function solution to $\psi$. However in the Kasner epoch that was discussed for large $\gamma$ and after collapse, $e^{-\chi/2}$ and those proportional to it cannot be dropped. If one sets $\psi = \sqrt{2} \beta \log r$ into the above, the functions that were required appear.

\printbibliography

\end{document}